\documentclass[vecphys]{svmult}

\usepackage{graphicx}        
\usepackage[bottom]{footmisc}
\usepackage{axodraw}
\usepackage{amssymb,amsmath}
\usepackage{pstricks,pst-plot}                                  
\newcommand{\downrightarrow}{\hspace{-3.3ex} \raisebox{0.2ex}{$\stackrel{ \!\! \mid }{\raisebox{-0.4ex}[0ex][0ex]{$ \;\;\;\;\;\; \longrightarrow \;\; $}}$}}
\newcommand{\lsim}{\mbox{\raisebox{-0.3ex}{\footnotesize $\:\stackrel{<}{\sim}\:$}} }

\begin{document}

\title*{Essentials of the Muon $g-2$}
\author{F. Jegerlehner\inst{1}}
\authorrunning{The Muon $g-2$}
\institute{Humboldt-Universit\"at zu Berlin, Institut f\"ur Physik,
       Newtonstrasse 15, D-12489 Berlin, Germany and DESY,
       Platanenallee 6, D-15738 Zeuthen, Germany
\texttt{fjeger@physik.hu-berlin.de}
}

\onecolumn{
\renewcommand{\thefootnote}{\fnsymbol{footnote}}
\setlength{\baselineskip}{0.52cm}
\thispagestyle{empty}
\begin{flushright} \begin{tabular}{c}
DESY 07-033 \\
HU-EP-07/05\\
March 2007\\
\end{tabular}
\end{flushright}
\hfill (update\footnote[4]{Including the new result on the universal
$O(\alpha^4)$ term of T.~Aoyama, M.~Hayakawa, T.~Kinoshita, M.~Nio,
arXiv:0706.3496 [hep-ph], which implies a 7 $\sigma$ shift in
$\alpha$. Note that with $\alpha$ defined via $a_e$ the change in the
universal part of $g-2$ only modifies the bookkeeping but does not
affect the final result as $a_e^{\rm uni}=a_\mu^{\rm uni}$ and the
non-universal part of $a_e$ only accounts for $(a_{e}^{\rm
exp}-a_{e}^{\rm uni})/a_{e}^{\rm exp}=3.8$ parts per billion} June 2007)
\setcounter{page}{0}

\mbox{}
\vspace*{\fill}
\begin{center}
{\Large\bf
Essentials of the Muon g-2}

\vspace{5em}
\large
F. Jegerlehner\footnote[1]{\noindent
Work supported in part by the European Community's Human Potential Program under
contract HPRN-CT-2002-00311 EURIDICE and the TARI Program under
contract RII3-CT-2004-506078.}
\\
\vspace{5em}
\normalsize
{\it Humboldt-Universit\"at zu Berlin, Institut f\"ur Physik,\\
       Newtonstrasse 15, D-12489 Berlin, Germany}\\

        and

{\it Deutsches Elektronen-Synchrotron DESY,\\
       Platanenallee 6, D-15738 Zeuthen, Germany}\\
\end{center}
\vspace*{\fill}}
\newpage

\maketitle

Abstract: The muon anomalous magnetic moment is one of the most
precisely measured quantities in particle physics. Recent high
precision measurements (0.54ppm) at Brookhaven reveal a ``discrepancy''
by 3.2 standard deviations from the electroweak Standard Model which
could be a hint for an unknown contribution from physics beyond the
Standard Model. This triggered numerous speculations about the
possible origin of the ``missing piece''. The remarkable 14-fold
improvement of the previous CERN experiment, actually animated a
multitude of new theoretical efforts which lead to a substantial
improvement of the prediction of $a_\mu$. The dominating uncertainty
of the prediction, caused by strong interaction effects, could be
reduced substantially, due to new hadronic cross section measurements
in electron-positron annihilation at low energies. After an
introduction and a brief description of the principle of the
experiment, I present a major update and review the status of the
theoretical prediction and discuss the role of the hadronic vacuum
polarization effects and the hadronic light--by--light scattering
contribution.  Prospects for the future will be briefly discussed. As,
in electroweak precision physics, the muon $g-2$ shows the largest
established deviation between theory and experiment at present, it
will remain one of the hot topics for further investigations.

\section{Lepton magnetic moments}
The subject of our interest is the motion of a lepton in an external
electromagnetic field under consideration of the full relativistic
quantum behavior.  The latter is controlled by the equations of motion
of Quantum Electrodynamics (QED), which describes the interaction of
charged leptons ($\ell=e,\mu,\tau$) with the photon ($\gamma$) as an
Abelian $U(1)_\mathrm{em}$ gauge theory. QED is a quantum field theory
(QFT) which emerges as a synthesis of quantum mechanics with special
relativity. In our case an external electromagnetic field is added,
specifically a constant homogeneous magnetic field $\vec{B}$. For
slowly varying fields the motion is essentially determined by the
generalized Pauli equation, which also serves as a basis for
understanding the role of the magnetic moment of a lepton on the
classical level. As we will see below, in the absence of electrical
fields $\vec{E}$ the quantum correction miraculously may be subsumed
in a single number, the anomalous magnetic moment $a_\ell$, which is
the result of relativistic quantum fluctuations, usually simply called
\textit{radiative corrections}.

Charged leptons in first place interact with photons, and photonic
radiative corrections can be calculated in QED, the interaction
Lagrangian density of which is given by ($e$ is the magnitude of the
electron's charge)
\begin{equation}
{\cal L}^{\mathrm{QED}}_{\mathrm{int}}(x)= e j^\mu_\mathrm{em}(x)\:A_\mu(x)\;\;,\;\;
j^\mu_\mathrm{em}(x)=-\sum_\ell \bar{\psi}_\ell(x)\gamma^\mu \psi_\ell(x)\; ,
\end{equation}
where $j^\mu_\mathrm{em}(x)$ is the
electromagnetic current,
$\psi_\ell(x)$ the Dirac field describing the lepton $\ell$, $\gamma^\mu$ the
Dirac matrices and with
a photon field $A_\mu(x)$ exhibiting an external
classical component $A^{\mathrm{ext}}_\mu$ and hence
$A_\mu \to A_\mu + A^{\mathrm{ext}}_\mu \;.$
We are thus dealing with QED exhibiting an additional
external field insertion ``vertex''.

Besides charge, spin, mass and lifetime, leptons have other very
interesting static (classical) electromagnetic and weak properties
like the magnetic and electric dipole moments. Classically the dipole
moments can arise from either electrical \textit{charges} or
\textit{currents}. A well known example is the circulating current, due to an
orbiting particle with electric charge $e$ and mass $m$, which
exhibits a magnetic dipole moment
$\vec{\mu}_L=\frac{e}{2c}\:\vec{r}\times \vec{v}$ given by
\begin{equation}
\vec{\mu}_L= \frac{e}{2mc}\:\vec{L}
\label{02Orbitalmm}
\end{equation}
where $\vec{L}=m\:\vec{r} \times \vec{v}$ is the orbital angular
momentum ($\vec{r}$ position, $\vec{v}$ velocity).
As we know, most elementary particles have intrinsic angular momentum,
called spin, and in particular leptons like the electron are Dirac fermions
of spin $\frac12$. Spin is directly responsible for the intrinsic
magnetic moment of any spinning particle.
The fundamental relation which defines the ``$g$--factor'' or the
magnetic moment is
\begin{equation}
\vec{\mu}={ g_\ell}\:\frac{e \hbar}{2m_\ell
c}\:\vec{S}\;,\;\;\vec{S}~~\mathrm{the \ spin \ vector.}
\end{equation}
For leptons, the Dirac theory predicts $g_\ell=2$~\cite{02Diracmm28},
unexpectedly, twice the value $g=1$ known to be associated with
orbital angular momentum.  It took about 20 years of experimental
efforts to establish that the electrons magnetic moment actually
exceeds 2 by about 0.12\%, the first clear indication of the existence
of an ``anomalous'' contribution to the magnetic moment~\cite{02Kusch48}.
In general, the anomalous magnetic moment of a lepton is related to the
gyromagnetic ratio by
\begin{equation}
a_\ell=\mu_\ell/\mu_B-1=\frac12 (g_\ell-2) \;,\;\; (\ell=e,\mu,\tau)
\label{02amudef}
\end{equation}
where $\mu_B$ is the Bohr magneton which has the value
\begin{equation}
\mu_B=\frac{e\hbar}{2m_ec}=5.788381804(39) \times 10^{-11}~~\mathrm{MeV}
\mathrm{T}^{-1} \;.
\end{equation}

Formally, the anomalous magnetic moment is given by a form factor,
defined by the matrix element $$\langle{\ell^-(p')}|
j_\mathrm{em}^\mu(0) |{\ell^-(p)\rangle}$$ where $|{\ell^-(p)\rangle}$
is a lepton state of momentum $p$. The relativistically covariant
decomposition of the matrix element reads

\vspace*{1.6cm}

\begin{picture}(120,60)(-10,-7)
\SetScale{2.0}
\Photon(00,30)(15,30){1}{4}
\ArrowLine(27.5,36)(36.5,45)
\ArrowLine(36.5,15)(27.5,24)
\Line(21.5,30)(36.5,45)
\Line(36.5,15)(21.5,30)
\put(10,76){\makebox(0,0)[t]{$\gamma(q)$}}
\put(83,86){\makebox(0,0)[t]{$\mu(p')$}}
\put(83,46){\makebox(0,0)[t]{$\mu(p)$}}
\GCirc(21.5,30){6.5}{0.75}
\end{picture}

\vspace*{-2.75cm}

$ \hspace{4cm} =(-\I e)\:\bar{u}(p')\left[\gamma^\mu
F_\mathrm{E}(q^2)+\I\frac{\sigma^{\mu\nu}q_\nu}{2m_\mu}F_\mathrm{M}(q^2) \right]u(p)$

\vspace*{1cm}

\noindent with $q=p'-p$ and
where $u(p)$ denotes a Dirac spinor, the relativistic wave function of a free
lepton, a classical solution of the Dirac equation $(\gamma^\mu p_\mu-m)\:u(p)=0$.
$F_\mathrm{E}(q^2)$ is the electric charge or Dirac form
factor and $F_\mathrm{M}(q^2)$ is the magnetic or Pauli
form factor. Note that the matrix $\sigma^{\mu
\nu}=\frac{\I}{2}[\gamma^\mu, \gamma^\nu]$ represents the spin $1/2$
angular momentum tensor. In the static (classical) limit $q^2 \to 0$ we have
\begin{equation}
F_\mathrm{E}(0)=1\;\;;\;\;\; F_\mathrm{M}(0)=a_\mu
\end{equation}
where the first relation is the charge normalization condition, which
must be satisfied by the electrical form factor, while the second
relation defines the anomalous magnetic moment. $a_\mu$ is a finite prediction
in any renormalizable QFT: QED, the Standard Model (SM) or
any renormalizable extension of it.

By end of the 1940's the
breakthrough in understanding and handling renormalization of QED
(Tomonaga, Schwinger, Feynman, and others) had
made unambiguous predictions of higher order effects possible, and in
particular of the leading (one-loop diagram) contribution to the
anomalous magnetic moment
\begin{equation}
a^{\mathrm{QED}(1)}_\ell = \frac{\alpha}{2\pi} \;,
(\ell=e,\mu,\tau)
\label{02Sch48}
\end{equation}
by Schwinger in 1948~\cite{02Sch48}. This contribution is due to quantum
fluctuations via virtual photon-lepton interactions and in QED is
universal for all leptons. At higher orders, in the perturbative
expansion\footnote{which is equivalent to the loop-expansion, referring
to the number of closed loops in corresponding Feynman diagrams.},
other effects come into play: strong interaction, weak interaction, both
included in the SM, as well as yet unknown physics which would contribute to
the anomalous magnetic moment.

In fact, shortly before Schwinger's QED prediction, Kusch and Foley in 1948
established the existence of the electron ``anomaly'' $g_e=2\:(1.00119\pm0.00005)$, a 1.2
per mill deviation from the value 2 predicted by Dirac in 1928.

We now turn to the muon.
A muon looks like a copy of an electron, which at first sight is just
much heavier $m_\mu/m_e\sim 200$, however, unlike the electron it is
unstable and its lifetime is actually rather short. The decay proceeds by
weak charged current interaction into an electron and two neutrinos.

The muon is very interesting for the following reason:
quantum fluctuations due to heavier particles or contributions from higher energy
scales are proportional to
\begin{equation}
\frac{\delta a_\ell}{a_\ell} \propto \frac{m_\ell^2}{M^2}~~~~~~~~(M \gg m_\ell)\;,
\end{equation}
where $M$ may be
\begin{itemize}
\item[~~-~~] the mass of a heavier SM particle, or
\item[~~-~~] the mass of a hypothetical heavy state beyond the SM, or
\item[~~-~~] an energy scale or an ultraviolet cut-off where the SM
ceases to be valid.
\end{itemize}
On the one hand, this means that the heavier the new state or scale
the harder it is to see (it decouples as $M \to \infty$). Typically
the best sensitivity we have for nearby new physics, which has not yet
been discovered by other experiments. On the other hand, the
sensitivity to ``new physics'' grows quadratically with the mass of
the lepton, which means that the interesting effects are magnified in
$a_\mu$ relative to $a_e$ by a factor $(m_\mu/m_e)^2
\sim 4 \times 10^{4}$. This is what makes the anomalous magnetic moment
of the muon the predestinated ``monitor for new physics'' or, if no
deviation is found it may provide severe constraints to physics beyond
the SM\footnote{Even more promising would be a measurement of $a_\tau$
with additional enhancement $(m_\tau/m_\mu)^2\sim 283$. However, the
much shorter lifetime of the $\tau$ lepton ($\tau_\tau/\tau_\mu \sim
1.3 \times 10^{-7}$) makes this measurement impossible
at present.}.

In contrast, $a_e$ is relatively insensitive to unknown physics and
can be predicted very precisely, and
therefore it presently provides the most precise determination of the
fine structure constant $\alpha=e^2/4\pi$.

What makes the muon so special for what concerns its  anomalous magnetic moment?
\begin{itemize}
\item Most interesting is the enhanced high sensitivity of $a_\mu$ to all kind of
interesting physics effects.
\item Both experimentally and theoretically $a_\mu$ is a ``clean''
observable, i.e., it can be measured with high precision as well as
predicted unambiguously in the SM.
\item That $a_\mu$ can be measured so precisely,  is kind of a miracle
and possible only due to the specific properties of the muon.
Due to the parity violating weak (V-A) interaction property, muons can easily
be polarized and perfectly transport polarization information to the electrons
produced in their decay.
\item There exists a magic energy (``magic $\gamma$'') at which
equations of motion take a particularly simple form. Miraculously,
this energy is so high (3.1 GeV) that the $\mu$ lives 30
times longer than in its rest frame!
\end{itemize}
In fact only these highly energetic muons can by collected in a muon
storage ring. At much lower energies muons could not be stored long
enough to measure the precession precisely!

\begin{figure}[t]
\centering
\includegraphics[height=8.7cm]{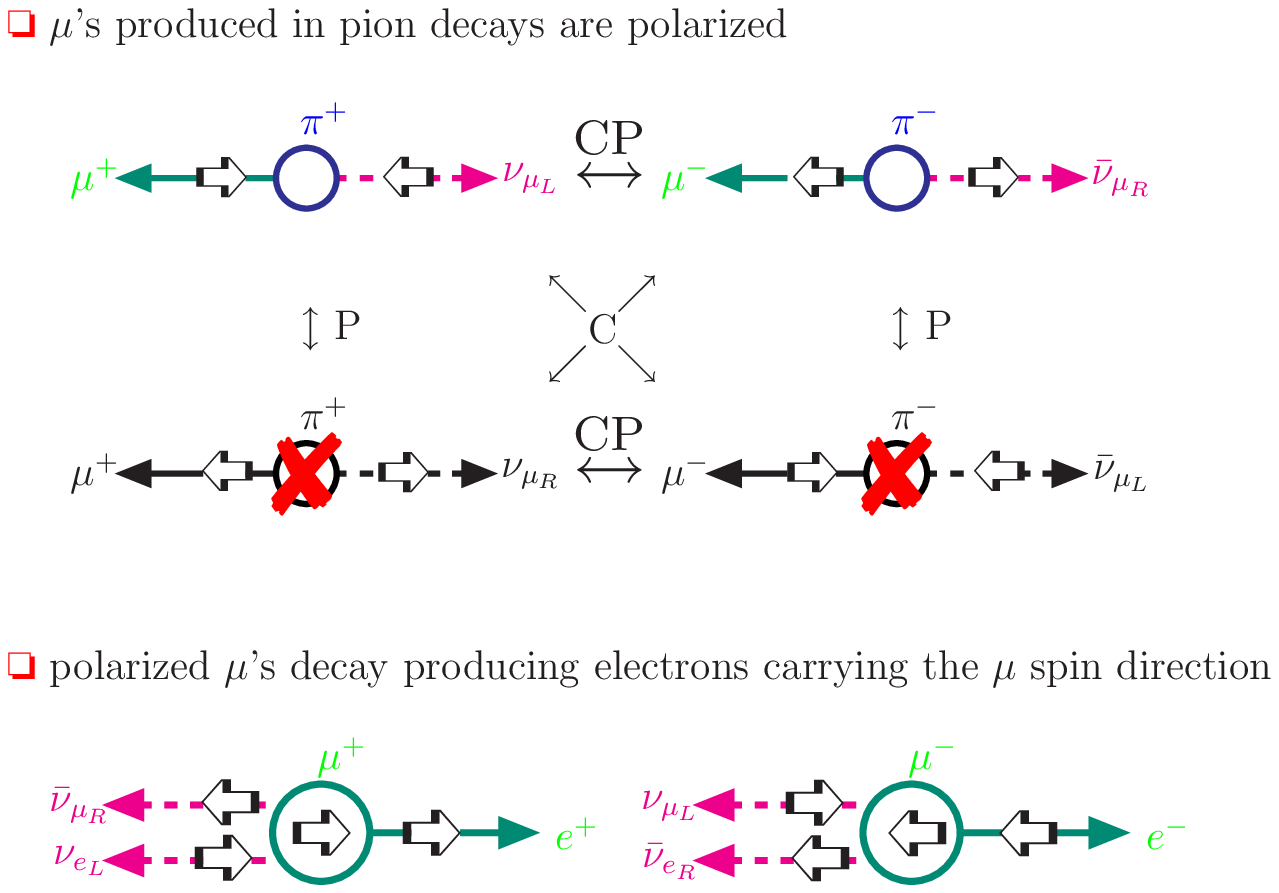}
\caption{Spin transfer properties in production and decay of the muons
(P=parity, C=charge conjugation). $\mu^-$ [$\mu^+$] is produced
with positive [negative]
helicity $h=\vec{s}\cdot\vec{p}/|\vec{p}|$, decay $e^-$ [$e^+$] have
negative [positive] helicity, respectively}
\label{02fig:muonprodec}
\vspace*{-2.3cm}
\begin{picture}(270,200)(0,-168)
\put(-37,10){\framebox(340,240)}
\end{picture}
\vspace*{-5.0cm}
\end{figure}

Production and decay of the muons goes by the chain
\begin{eqnarray*}
\pi &\to& \mu + \nu_\mu \\
 && \downrightarrow e + \nu_e + \nu_\mu
\end{eqnarray*}
and the polarization ``gymnastics'' is illustrated in Fig.~\ref{02fig:muonprodec}.
Note that the ``maximal'' parity ($P$) violation means that the
charged weak transition currents only couple to left-handed neutrinos
$\nu_{\mu \raisebox{-0.3ex}{\tiny $L$}}$ and right-handed
antineutrinos $\bar{\nu}_{\mu
\raisebox{-0.3ex}{\tiny $R$}}$, in other words, parity
violation is a direct consequence of the fact that the neutrinos
$\nu_{\mu \raisebox{-0.3ex}{\tiny $R$}}$ and $\bar{\nu}_{\mu
\raisebox{-0.3ex}{\tiny $L$}}$ show no electromagnetic, weak and  strong
interaction in nature! as if they were non-existent.

\section{The BNL Muon $g-2$ Experiment}
After the proposal of parity violation in weak transitions by Lee and
Yang in 1957, it immediately was realized that muons produced in weak
decays of the pion ($\pi^+ \to \mu^+ + $ neutrino) should be
longitudinally polarized. In addition, the decay positron of the muon
($\mu^+ \to e^++2 $ neutrinos) could indicate the muon spin
direction. This was confirmed by Garwin, Lederman and
Weinrich~\cite{02Garwin57} and Friedman and
Telegdi~\cite{02Friedman57}\footnote{The latter reference for the
first time points out that $P$ and $C$ are violated simultaneously, in
fact $P$ is maximally violated while $CP$ is to very good
approximation conserved in this decay (see Fig.~\ref{02fig:muonprodec}).}. The first of the two papers
for the first time determined $g_\mu=2.00$ within 10\% by applying the
muon spin precession principle. Now the road was free to seriously
think about the experimental investigation of $a_\mu$.

The first measurement of $(g_\mu-2)/2$ was performed at Columbia in
1960~\cite{02Columbia60} with a result $a_\mu=0.00122(8)$ at a
precsision of about 5\%. Soon later in 1961, at the CERN cyclotron
(1958-1962) the first precision determination became
available~\cite{02CERN62}. Surprisingly, nothing special was observed
within the 0.4\% level of accuracy of the experiment. It was the first
real evidence that the muon was just a heavy electron. In particular
this meant that the muon is point-like and no extra short distance
effects could be seen. This latter point of course is a matter of
accuracy and the challenge to go further was evident.

The idea of a muon storage ring was put forward next.  A first one was
successfully realized at CERN (1962-1968)~\cite{02CERN66}.  It allowed
to measure $a_\mu$ for both $\mu^+$ and $\mu^-$ at the same
machine. Results agreed well within errors and provided a precise
verification of the CPT theorem for muons. An accuracy of 270 ppm was
reached and an insignificant 1.7 $\sigma$ (1 $\sigma$ = 1 standard
deviation) deviation from theory was found. Nevertheless the latter
triggered a reconsideration of theory. It turned out that in the
estimate of the three-loop $O(\alpha^3)$ QED contribution the leptonic
``light-by-light scattering'' part in the radiative corrections
(dominated by the electron loop) was missing. Aldins et
al.~\cite{02LBLlep69} then calculated this and after including it,
perfect agreement between theory and experiment was obtained.\\
\begin{figure}
\centering
\includegraphics[height=7.2cm]{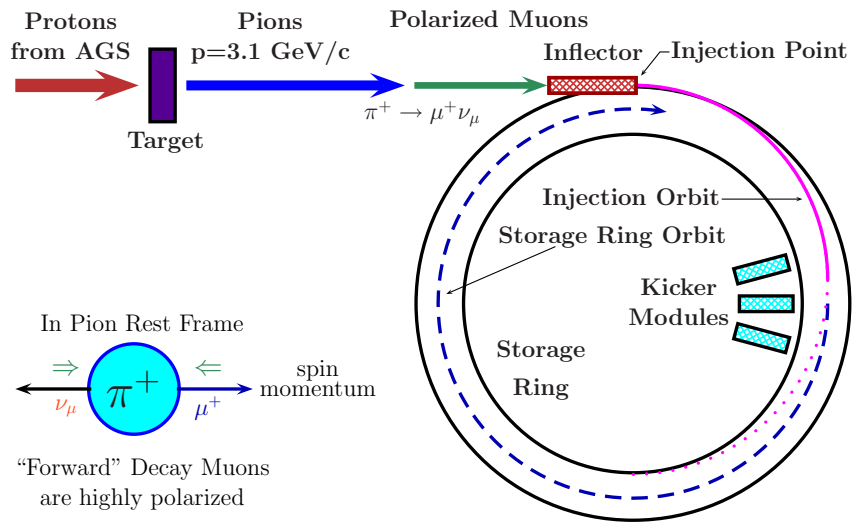}
\caption{The schematics of muon injection and storage in the $g-2$ ring}
\label{02fig:SRschema}
\end{figure}

\vspace*{-3mm}

The CERN muon $g-2$ experiment was shut down end of 1976, while data
analysis continued until 1979~\cite{02Bailey79}. Only a few years
later, in 1984 the E821collaboration formed, with the aim to perform a
new experiment at Brookhaven National Laboratory (BNL). Data taking
was between 1998 and 2001. The data analysis was completed in 2004.
The E821 $g-2$ measurements achieved the remarkable precision of
0.54ppm~\cite{02BNL04,02BNLfinal}, which is a 14-fold improvement of
the CERN result.  The principle of the BNL muon $g-2$ experiments
involves the study of the orbital and spin motion of highly polarized
muons in a magnetic storage ring. This method has been applied in the
last CERN experiment already. The key improvements of the BNL
experiment include the very high intensity of the primary proton beam
from the Alternating Gradient Synchrotron (AGS), the injection of
muons instead of pions into the storage ring, and a superferric
storage ring magnet. The protons hit a target and produce pions. The
pions are unstable and decay into muons plus a neutrino where the
muons carry spin and thus a magnetic moment which is directed along
the direction of the flight axis. The longitudinally polarized muons
from pion decay are then injected into a uniform magnetic field
$\vec{B}$ where they travel in a circle (see
Fig.~\ref{02fig:SRschema}).\\[-18mm]

\begin{figure}
\centering
\includegraphics[height=8.6cm]{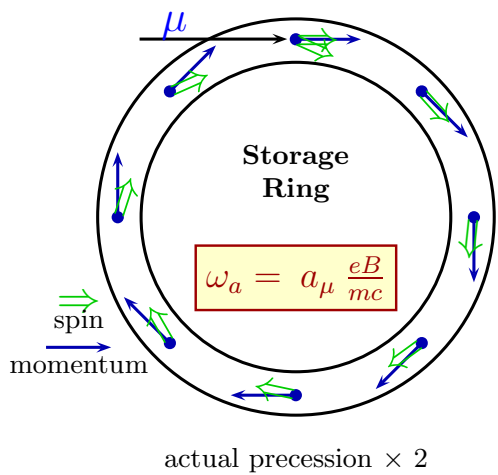}
\vspace*{-9mm}
\caption{Spin precession in the $g-2$ ring ($\sim 12^\circ$/circle)}
\label{02fig:spinprecession}
\end{figure}

\vspace*{-3mm}

When polarized muons travel on a circular orbit in a constant
magnetic field, as illustrated in Fig.~\ref{02fig:spinprecession},
then $a_\mu$ is responsible for the Larmor precession of the direction
of the spin of the muon, characterized by the angular fequency
$\vec{\omega}_a$. At the magic energy of about $\sim$ 3.1 GeV, the
latter is directly proportional to $a_\mu$:\\[-2mm]
\begin{equation}
\vec{\omega}_a=\frac{e}{m}\left[
a_\mu \vec{B}-\left(a_\mu-\frac{1}{\gamma^2-1} \right)\:
\vec{\beta} \times \vec{E}
 \right]_{{\mathrm{at}\;"\mathrm{magic} \ \gamma"}}^{ E \sim 3.1 \mathrm{GeV} }
\simeq \frac{e}{m}\left[
a_\mu \vec{B} \right]\;.
\label{02measuringamu}
\end{equation}
Electric quadrupole fields $\vec{E}$ are needed for focusing the beam
and they affect the precession frequency in
general. $\gamma=E/m_\mu=1/\sqrt{1-\beta^2}$ is the relativistic
Lorentz factor with $\beta=v/c$ the velocity of the muon in units of
the speed of light $c$. The magic energy
$E_\mathrm{mag}=\gamma_\mathrm{mag}m_\mu$ is the energy $E$ for which
$\frac{1}{\gamma_\mathrm{mag}^2-1}= a_\mu$.  The existence of a
solution is due to the fact that $a_\mu$ is a positive constant in
competition with an energy dependent factor of opposite sign (as
$\gamma \geq 1$). The second miracle, which is crucial for the
feasibility of the experiment, is the fact that
$\gamma_\mathrm{mag}=\sqrt{(1+a_\mu)/a_\mu}\simeq 29.378$ is large
enough to provide the time dilatation factor for the unstable muon
boosting the life time $\tau_\mu \simeq 2.197 \times 10^{-6}~
\mathrm{sec}$ to $\tau_\mathrm{in\: flight}=\gamma\: \tau_\mu
\simeq 6.454 \times 10^{-5}~
\mathrm{sec}$, which allows the muons, traveling at
$v/c=0.99942\cdots$, to be stored in a ring of reasonable size
(diameter $\sim$
14 m).

\begin{figure}[t]
\centering
\vspace*{-1cm}
\includegraphics[height=8cm]{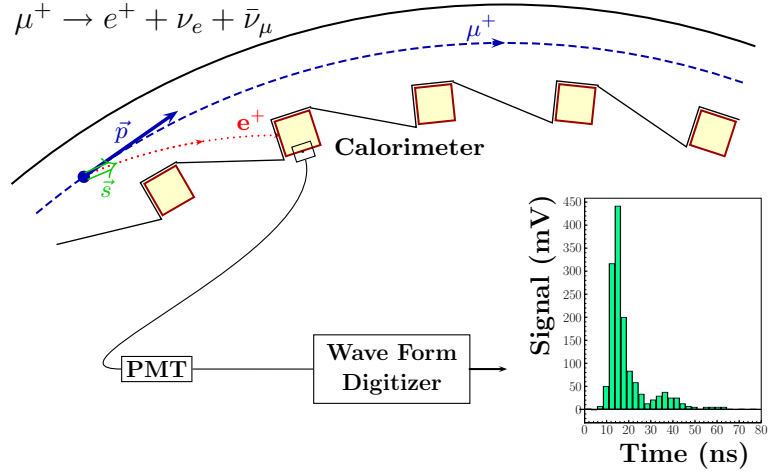}
\caption{Decay of $\mu^+$ and detection of the emitted $e^+$
(PMT=Photomultiplier)}
\label{02fig:detection}
\end{figure}

\begin{figure}[t]
\centering
\includegraphics[height=8cm]{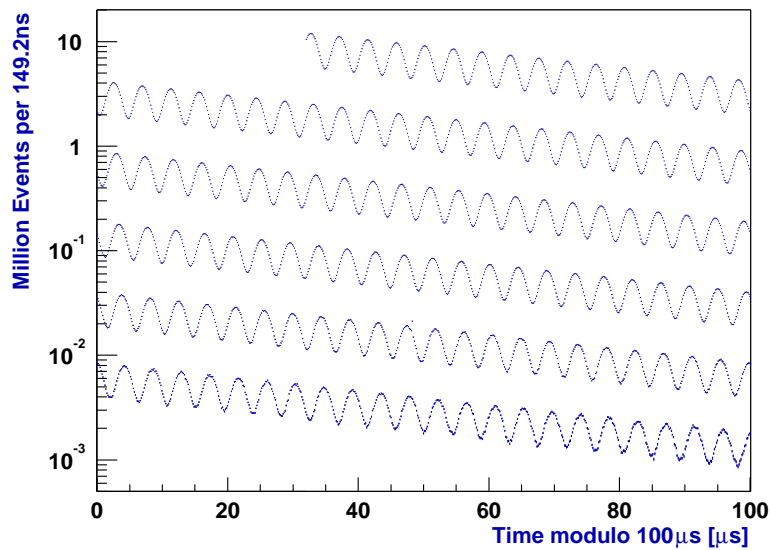}
\caption{Distribution of counts
versus time for the 3.6 billion decays in the 2001 negative muon
data-taking period. Courtesy of the E821 collaboration~\cite{02BNL04}}
\label{02fig:wiggle2004}
\end{figure}

\begin{figure}[t]
\centering
\includegraphics[width=11.5cm,clip=true]{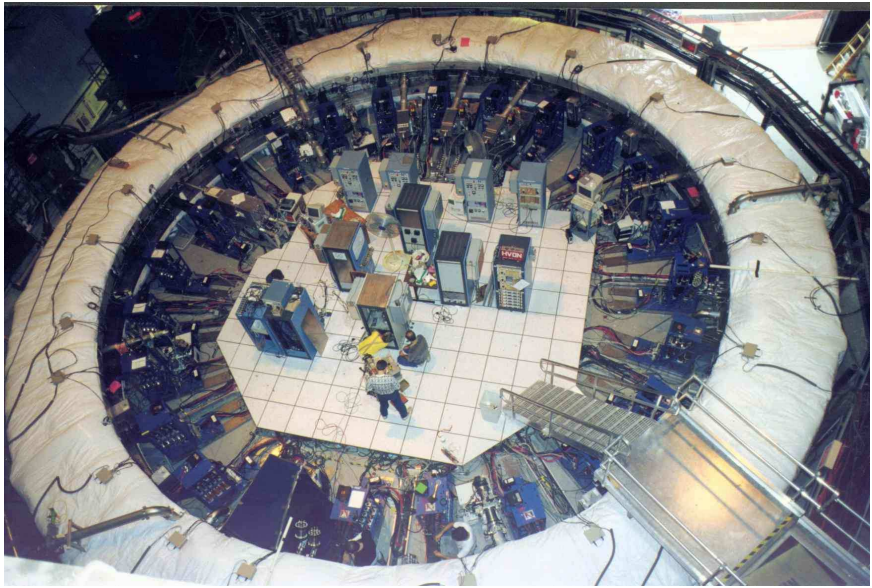}
\caption{The Brookhaven National Laboratory muon storage ring. The ring has
a radius of 7.112 meters, the aperture of the beam pipe is 90 mm,
the field is 1.45 Tesla and the momentum of the muon is $p_\mu =3.094$
GeV/c. Picture taken from the Muon $g-2$ Collaboration Web Page {\tt
http://www.g-2.bnl.gov/} (Courtesy of Brookhaven National Laboratory)}
\label{02fig:BNLring}
\end{figure}

This provided the basic setup for the $g-2$ experiments at the muon
storage rings at CERN and at BNL. The oscillation frequency
$\vec{\omega}_a$ can be measured very precisely. Also the precise
tuning to the magic energy is not the major problem.  The most serious
challenge is to manufacture a precisely known constant magnetic field
$B$, as the latter directly enters the experimental extraction of
$a_\mu$ (\ref{02measuringamu}). Of course one also needs high enough
statistics to get sharp values for the oscillation frequency. The
basic principle of the measurement of $a_\mu$ is a measurement of the
``anomalous'' frequency difference
$\omega_a=|\vec{\omega}_a|=\omega_s-\omega_c$, where
$\omega_s=g_\mu\:(e\hbar/2m_\mu)\:B/\hbar=g_\mu/2\:\cdot ~ e/m_\mu
\:B$ is the muon spin--flip precession frequency in the applied
magnetic field and $\omega_c=e/m_\mu\:B$ is the muon cyclotron
frequency. The principle of measuring $\omega_a$ is indicated in
Fig.~\ref{02fig:detection} and an example of a measured count spectrum
is shown in Fig.~\ref{02fig:wiggle2004}. Instead of eliminating the
magnetic field by measuring $\omega_c$, $B$ is determined from proton
nuclear-magnetic-resonance (NMR) measurements. This procedure requires
the value of $\mu_\mu/\mu_p$ to extract $a_\mu$ from the
data. Fortunately, a high precision value for this ratio is available
from the measurement of the hyperfine structure in muonium. One
obtains\\[-2mm]
\begin{equation}
a_\mu=\frac{\bar{R}}{|\mu_\mu/\mu_p|-\bar{R}}\;,
\end{equation}
where $\bar{R}=\omega_a/\bar{\omega}_p$ and
$\bar{\omega}_p=(e/m_\mu c)\langle B \rangle$ is the
free-proton NMR frequency corresponding to the
average magnetic field, seen by the muons in their orbits in the
storage ring. We mention that for the electron a Penning trap is
employed to measure $a_e$ rather than a storage ring. The $B$ field in
this case can be eliminated via a measurement of the cyclotron
frequency. The BNL g-2 muon storage ring is shown in
Fig.~\ref{02fig:BNLring}.

Since the spin precession frequency can be measured very well, the
precision at which $g-2$ can be measured is essentially determined by
the possibility to manufacture a constant homogeneous magnetic field
$\vec{B}$. Important but easier to achieve is the tuning to the magic
energy. The outcome of the experiment will be discussed later.

\section{QED Prediction of $a_e$ and the Determination
of $\alpha$}

The anomalous magnetic moment $a_\ell$ is a dimensionless quantity,
just a number, and corresponds to an effective tensor interaction term
\begin{equation}
\delta {\cal L}_{\mathrm{eff}}^{\mathrm{AMM}}=
 - \frac{e_\ell a_\ell}{4m_\ell}\: \bar{\psi} (x) \:\sigma^{\mu \nu}\: F_{\mu
\nu} (x) \:\psi (x)  \;,
\label{02AMMeff}
\end{equation}
which in an external magnetic field at
low energy takes the well known form of a magnetic energy (up to a sign)
\begin{equation}
\delta {\cal L}_{\mathrm{eff}}^{\mathrm{AMM}} \Rightarrow  -{\cal H}_m \simeq
 -  \frac{e_\ell a_\ell}{2m_\ell} \:
 \vec{\sigma} \vec{B} \;.
\end{equation}
Such a term, if present in the fundamental Lagrangian, would spoil
renormalizability of the theory and contribute to $F_{\rm M}(q^2)$ at
the tree level. In addition, it is not $SU(2)_L$ gauge invariant,
because gauge invariance only allows minimal couplings via a covariant
derivative, i.e., vector and/or axial-vector terms. The emergence of
an anomalous magnetic moment term in the SM is a consequence of the
symmetry breaking by the Higgs mechanism, which provides the mass to
the physical particles and allows for helicity flip processes like the
anomalous magnetic moment transitions. In any renormalizable theory
the anomalous magnetic moment term must vanish at tree level. This
means that there is no free adjustable parameter associated with
it. It is a finite prediction of the theory.

The reason why it is so interesting to have such a precise measurement
of $a_e$ or $a_\mu$, of course, is that it can be calculated with
comparable accuracy in theory by a perturbative expansion in $\alpha$
of the form
\begin{equation}
a_\ell\simeq\sum_{n=1}^N A^{(2n)} (\alpha/\pi)^n \;,
\label{02aeexpansion}
\end{equation}
with up to $N=5$ terms under consideration at present.
The experimental precision of $a_e$ (0.66 ppb) requires the knowledge of
the coefficients with accuracies $\delta A^{(4)}\sim 1 \times 10^{-7}$,
$\delta A^{(6)}\sim 6 \times 10^{-5}$, $\delta A^{(8)}\sim 2 \times 10^{-2}$ and
$\delta A^{(10)}\sim 10$. The expansion (\ref{02aeexpansion}) is an expansion in
the number $N$ of closed loops of the contributing Feynman diagrams.

The recent new determination of $a_e$~\cite{02aenew} allows for a very precise
determination of the fine structure constant~\cite{02alnew,02Aoyama07}
\begin{eqnarray}
\alpha^{-1}(a_e)&=&137.035999069(96)[0.70 \mathrm{ppb}]\;,
\label{02alphainv_a_e}
\end{eqnarray}
which we will use in the evaluation of $a_\mu$.

At two and more loops results depend on lepton mass ratios. For
the evaluation of these contributions precise values for the lepton
masses are needed. We will use the following values for the
muon--electron mass ratio, the muon and the tau
mass~\cite{02PDG04,02CODATA02}
\begin{eqnarray} \begin{array}{c}
m_{\mu}/m_e =  206.768\,2838\,(54) ~,~~
m_{\mu}/m_\tau =  0.059\,4592\,(97) \\
m_e     =  0.510\,9989\,918(44) \mathrm{MeV} ~,~~
m_{\mu}     =  105.658\,3692\,(94) \mathrm{MeV} \\
m_{\tau}     = 1776.99\,(29) \mathrm{MeV}\;.
\end{array}
\label{02leptonmasses}
\end{eqnarray}

The leading contributions to $a_\ell$ can be calculated in QED. With
increasing precision higher and higher terms become relevant. At
present, 4--loops are indispensable and strong interaction effects
like hadronic vacuum polarization
(vap) or hadronic light-by-light scattering (lbl) as well as weak effects have to be
considered.  Typically, analytic results for higher order terms may be
expressed in terms of the Riemann zeta function
\begin{equation}
\zeta(n)= \sum\limits_{k=1}^{\infty} \frac{1}{k^n}
\label{02Zetadef}
\end{equation}
and of the poly-logarithmic integrals
\begin{equation}
\mathrm{Li}_n(x)=
\frac{(-1)^{n-1}}{(n-2)!}\int\limits_{0}^{1}\frac{\ln^{n-2}(t)\ln(1-tx)}{t}dt=
\sum\limits_{k=1}^{\infty}\frac{x^k}{k^n}\;.
\label{02Lindef}
\end{equation}

We first discuss the universal contributions $a_\ell$ in ``one
flavor QED'', with one type of
lepton lines only. At leading order
one has \\
$ \bullet$ one 1-loop diagram\\

\SetWidth{1}
\begin{picture}(30,30)(-40,-10)
\Line(00,00)(-21.213,-21.213)
\Line(00,00)(+21.213,-21.213)
\Photon(00,00)(00,16){1}{4}
\Photon(-10.607,-10.607)(+10.607,-10.607){1}{5}
\Text(-15,-25)[]{$ \ell$}
\Text(15,-25)[]{$ \ell$}
\Text(00,-20)[]{$ \gamma$}
\Text(160,-10)[]{$a_e=a_\mu=a_\tau=\frac{\alpha}{2\pi}$ \hspace{2cm} { Schwinger 48}}
\end{picture}

\vspace*{8mm}

\noindent
giving the result mentioned before.\\

$ \bullet$ At 2-loops 7 diagrams with only $\ell$-type fermion
lines\label{02twoloopuni}

\vspace*{1cm}

\SetScale{0.75}
\SetWidth{1}
\begin{picture}(30,30)(-20,-40)
\Line(00,00)(-21.213,-21.213)
\Line(00,00)(+21.213,-21.213)
\Photon(00,00)(00,16){1}{4}
\Photon(-07.071,-07.071)(+07.071,-07.071){1}{3}
\Photon(-14.142,-14.142)(+14.142,-14.142){1}{5}
\end{picture}
\begin{picture}(30,30)(-30,-40)
\Line(00,00)(-21.213,-21.213)
\Line(00,00)(+21.213,-21.213)
\Photon(00,00)(00,16){1}{4}
\Photon(-07.071,-07.071)(+14.142,-14.142){1}{5}
\Photon(-14.142,-14.142)(+07.071,-07.071){1}{5}
\end{picture}
\begin{picture}(30,30)(-40,-40)
\Line(00,00)(-21.213,-21.213)
\Line(00,00)(+21.213,-21.213)
\Photon(00,00)(00,16){1}{4}
\Photon(-10.607,-10.607)(+10.607,-10.607){1}{5}
\PhotonArc(-10.607,-10.607)(4.5,45,225){1}{4.5}
\end{picture}
\begin{picture}(30,30)(-50,-40)
\Line(00,00)(-21.213,-21.213)
\Line(00,00)(+21.213,-21.213)
\Photon(00,00)(00,16){1}{4}
\Photon(-10.607,-10.607)(+10.607,-10.607){1}{5}
\PhotonArc(+10.607,-10.607)(4.5,-45,135){1}{4.5}
\end{picture}
\begin{picture}(30,30)(-60,-40)
\Line(00,00)(-21.213,-21.213)
\Line(00,00)(+21.213,-21.213)
\Photon(00,00)(00,16){1}{4}
\PhotonArc(-07.071,-07.071)(4.5,45,225){1}{4.5}
\Photon(-14.142,-14.142)(+14.142,-14.142){1}{5}
\end{picture}
\begin{picture}(30,30)(-70,-40)
\Line(00,00)(-21.213,-21.213)
\Line(00,00)(+21.213,-21.213)
\Photon(00,00)(00,16){1}{4}
\PhotonArc(+07.071,-07.071)(4.5,-45,135){1}{4.5}
\Photon(-14.142,-14.142)(+14.142,-14.142){1}{5}
\end{picture}
\begin{picture}(30,30)(-80,-40)
\Line(00,00)(-21.213,-21.213)
\Line(00,00)(+21.213,-21.213)
\Photon(00,00)(00,16){1}{4}
\Photon(-14.142,-14.142)(+14.142,-14.142){1}{5}
\COval(00,-14.142)(3.536,5.000)(0){Black}{White}
\end{picture}

\vspace*{-5mm}
\noindent which contribute a term
\begin{equation}
a_\ell^{(4)}=\left[ \frac{197}{144}+ \frac{\pi^2}{12}-
\frac{\pi^2}{2}\ln 2 + \frac{3}{4} \zeta(3)\right] \: \left(\frac{\alpha}{\pi} \right)^2\;,
\end{equation}
obtained independently by Peterman~\cite{02Petermann57} and
Sommerfield~\cite{02Sommerfield57} in 1957.

\begin{figure}[t]
\vspace*{-7mm}
\centering \hspace*{-7mm}
\includegraphics[height=0.67\textwidth]{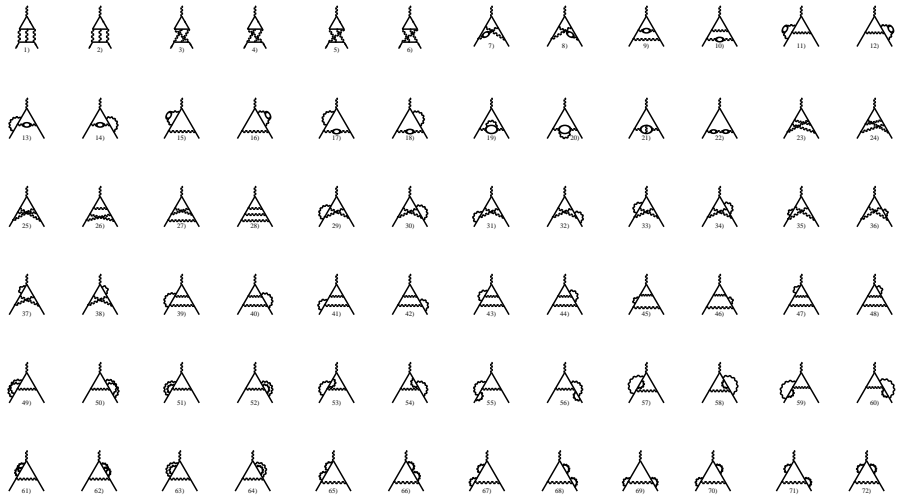}
\vspace*{-6mm}
\caption{The universal third order contribution to $a_\mu$. All
fermion loops here are muon--loops (first 22 diagrams). All
non-universal contributions follow by replacing at least one muon in a
closed loop by some other fermion}
\label{02fig:threeloopdiag}
\end{figure}

$ \bullet$ At 3-loops, with one type of fermion lines only, the 72
diagrams of Fig.~\ref{02fig:threeloopdiag} contribute. Most remarkably,
after about 25 years of hard work, Laporta and Remiddi in
1996~\cite{02LaportaRemiddi96} managed to give a complete analytic
result (see also~\cite{02Remiddietal69-95})
\begin{eqnarray}
a_\ell^{(6)}&=&\left[ \frac{28259}{5184}+ \frac{17101}{810} \pi^2 -
\frac{298}{9}\pi^2 \ln 2 + \frac{139}{18} \zeta(3)
+\frac{100}{3} \left\{ \mathrm{Li}_4(\frac{1}{2})
+\frac{1}{24}\ln^42 \right. \right. \nonumber \\ && \left. \left. -\frac{1}{24}\pi^2 \ln^22 \right\}
-\frac{239}{2160}\pi^4 + \frac{83}{72}\pi^2 \zeta(3)-\frac{215}{24}\zeta(5)
\right] \: \left(\frac{\alpha}{\pi} \right)^3 \;.
\end{eqnarray}
This result was confirming Kinoshita's earlier numerical
evaluation~\cite{02Ki95}.

The big advantage of the analytic result is that it allows a numerical
evaluation at any desired precision. The direct numerical
evaluation of the multidimensional Feynman integrals by Monte Carlo methods
is always of limited precision and an improvement  is always very expensive in
computing power.

$ \bullet$ At 4-loops 891 diagrams contribute to the universal term.
Their evaluation is possible by numerical integration and has been
performed in a heroic effort by Kinoshita~\cite{02KiLi83-89} (reviewed
in~\cite{02HuKi99}), and was updated recently by Kinoshita and his collaborators
(2002/2005/2007)~\cite{02KiNi05,02Aoyama07}.

The largest uncertainty comes from 518 diagrams without fermion loops
contributing to the universal term $A^{(8)}_1$. Completely unknown is
the universal five--loop term $A^{(10)}_1$, which is leading for
$a_e$. An estimation discussed in~\cite{02CODATA05} suggests that the
5-loop coefficient has at most a magnitude of 3.8. We adopt this
estimate and take into account $A^{(10)}_1 =0.0(3.8)$ (as
in~\cite{02KiNi05}).

Collecting the universal terms we have
\begin{eqnarray}
a_\ell^{\rm uni}&=&0.5 \: \left(\frac{\alpha}{\pi} \right)
-0.32847896557919378\ldots  \: \left(\frac{\alpha}{\pi} \right)^2
\nonumber \\&&
+1.181241456587 \ldots  \: \left(\frac{\alpha}{\pi} \right)^3
-1.9144(35) \: \left(\frac{\alpha}{\pi} \right)^4
+0.0(3.8) \: \left(\frac{\alpha}{\pi} \right)^5
\nonumber \\&=&
0.001\,159\,652\,176\,42(81)(10)(26)[86] \cdots
\end{eqnarray}
for the one--flavor QED contribution. The three errors are: the
error of $\alpha$, given in (\ref{02alphainv_a_e}), the numerical
uncertainty of the $\alpha^4$ coefficient and the estimated size of
the missing higher order terms, respectively.

At two loops and higher, internal fermion-loops show up, where the
flavor of the internal fermion differs form the one of the external
lepton, in general. As all fermions have different masses, the
fermion-loops give rise to mass dependent effects, which were
calculated at two-loops in~\cite{02SWP57,02El66} (see
also~\cite{02LdeR69,02Lautrup77,02LdeR74,02LiMeSa93}), and
at three-loops in~\cite{02Ki67,02A26early,02SaLi91,02La93,02LR93,02KOPV03}.

The leading mass dependent effects come from photon vacuum
polarization, which leads to charge screening.
Including a factor $e^2$ and considering the renormalized photon propagator
(wave function renormalization factor $Z_\gamma$) we have
\begin{equation}
\I \:e^2\:D^{' \mu \nu}_\gamma (q)=\frac{- \I g^{\mu \nu} \: e^2\:Z_\gamma}{q^2\:
\left( 1+\Pi'_\gamma(q^2)\right)}+ {\rm \
gauge \ terms \ }
\end{equation}
which in effect means that the charge has to be replaced by an
energy-momentum scale dependent {\em running charge}
\begin{equation}
e^2 \to e^2(q^2)=\frac{e^2Z_\gamma}{1+\Pi'_\gamma(q^2)} \;.
\end{equation}
The wave function renormalization factor $Z_\gamma$ is fixed by the
condition that as $q^2 \to 0$ one obtains the classical charge (charge
renormalization in the Thomson limit). Thus the renormalized charge is
\begin{eqnarray}
e^2 \to e^2(q^2)=\frac{e^2}{1+(\Pi'_\gamma(q^2)-\Pi'_\gamma(0))}
\label{02runninge}
\end{eqnarray}
where in perturbation theory the lowest order diagram which contributes
to $\Pi'_\gamma(q^2)$ is\\[-13mm]
\begin{figure}[h]
\centering
\includegraphics[height=1.1cm]{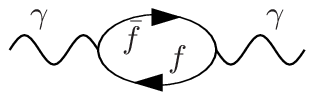}
\end{figure}

\vspace*{-7mm}

\noindent
and describes the virtual creation and re-absorption of fermion pairs
$\gamma^* \rightarrow e^+e^-,\mu^+ \mu^-, \tau^+ \tau^-, u\bar{u}, d
\bar{d},$ $\cdots (\mathrm{had})\rightarrow ~~\gamma^*$~.

In terms of the fine structure constant $\alpha=\frac{e^2}{4\pi}$ Eq.~({\ref{02runninge})
reads
\begin{equation}
\alpha(q^2)=\frac{\alpha}{1-\Delta \alpha(q^2)}\;\;\;;\;\;\;\Delta \alpha(q^2) =
- {\rm Re}\:\left(\Pi'_\gamma(q^2)-\Pi'_\gamma(0)\right) \;.
\label{02dalpdef}
\end{equation}
The various contributions to the shift in the fine structure constant
come from the leptons (lep = $e$, $\mu$ and $\tau$), the 5 light
quarks ($u$, $b$, $s$, $c$, and $b$) and/or the
corresponding hadrons (had). The top quark is too heavy
to give a relevant contribution. The hadronic contributions will be
considered later. The running of $\alpha$ is governed by the
renormalization group (RG). In the
context of $g-2$ calculations, the use of RG methods has been
advocated in~\cite{02LdeR74}. In fact, the enhanced short-distance
logarithms may be obtained by the substitution $\alpha \to
\alpha(m_\mu)=\alpha\:(1+\frac23 \frac{\alpha}{\pi}\:\ln \frac{m_\mu}{m_e}+\cdots)$ in a
lower order result (see the following example).

Typical contributions are the following: \\
 $-$ LIGHT internal masses give rise to log's of mass ratios which
become singular in the light mass to zero limit (logarithmically
enhanced corrections)\\[-3mm]
\SetWidth{2}
\begin{picture}(30,30)(-40,00)
\Line(00,00)(-21.213,-21.213)
\Line(00,00)(+21.213,-21.213)
\SetWidth{1}
\Photon(00,00)(00,16){1}{4}
\Photon(-14.142,-14.142)(+14.142,-14.142){1}{5}
\COval(00,-14.142)(3.536,5.000)(0){Black}{White}
\Text(-24,-19)[]{$ \mu$}
\Text(06,-19)[]{$ e$}
\Text(120,0)[]{$=\left[\frac{1}{3} \ln \frac{m_\mu}{m_e}-\frac{25}{36} + O\left(
\frac{m_e}{m_\mu}\right) \right]\:\left(
\frac{\alpha}{\pi}\right)^2\,.$}
\end{picture}

\vspace*{10mm}

$-$ HEAVY internal masses decouple, i.e., they give no effect in the
heavy mass to infinity limit \\[-3mm]
\SetWidth{1}
\begin{picture}(30,30)(-40,00)
\Line(00,00)(-21.213,-21.213)
\Line(00,00)(+21.213,-21.213)
\Photon(00,00)(00,16){1}{4}
\Photon(-14.142,-14.142)(+14.142,-14.142){1}{5}
\SetWidth{2}
\COval(00,-14.142)(3.536,5.000)(0){Black}{White}
\SetWidth{1}
\Text(-24,-19)[]{$ e$}
\Text(06,-19)[]{$ \mu$}
\Text(120,0)[]{$=\left[\frac{1}{45}\left(\frac{m_e}{m_\mu} \right)^2+ O\left(
\frac{m_e^4}{m_\mu^4}\ln \frac{m_\mu}{m_e} \right) \right]\:\left(
\frac{\alpha}{\pi}\right)^2\,.$}
\end{picture}

\vspace*{10mm}

New physics contributions  from states which are too heavy to be produced
at present accelerator energies typically give this kind of contribution.
Even so $a_\mu$ is 786 times less precise than $a_e$
it is still 54 times more sensitive to new physics (NP).

Corrections due to internal $e$, $ \mu$- and $ \tau$-loops are
different for $a_e$, $a_\mu$ and $a_\tau$. For reasons of comparison
and because of its role in the precise determination of $\alpha$ we
briefly consider $a_e$ first. The result is of the form
\begin{equation}
a_e^\mathrm{QED}=a_e^{\rm uni}+a_e(\mu)+a_e(\tau)+e_e(\mu,\tau)
\end{equation}
with\footnote{The order $\alpha^3$ terms are given by two parts
which cancel partly
\begin{eqnarray*}
A_2^{(6)}(m_e/m_\mu)
&=&\left. -2.17684015(11) \times 10^{-5}\right|_{\mu - \mathrm{vap}} +
\left. 1.439445989(77) \times
10^{-5}\right|_{\mu - \mathrm{lbl}} \\
A_2^{(6)}(m_e/m_\tau) &=&
\left. -1.16723(36) \times
10^{-7}\right|_{\tau - \mathrm{vap}} + \left. 0.50905(17) \times
10^{-7}\right|_{\tau - \mathrm{lbl}}\;.
\end{eqnarray*}
The errors are due to the uncertainties in the mass ratios. They are
negligible in comparison with the other errors. ``vap'' denotes
vacuum polarization type contributions~\cite{02La93} and ``lbl'' light-by-light
scattering type ones~\cite{02LR93} (the first 6 diagrams of
Fig.~\ref{02fig:threeloopdiag} with an $e$-- or $\tau$-- loop).}~\cite{02El66,02La93,02LR93}
\begin{eqnarray}
a_e(\mu)&=&5.197\,386\,70(27) \times 10^{-7}
{\left(\frac{\alpha}{\pi} \right)^2}
-7.373\,941\,64(29) \times 10^{-6} {\left(\frac{\alpha}{\pi}
\right)^3} \nonumber
\end{eqnarray}
\begin{eqnarray}
a_e(\tau)&=&1.83763(60) \times
10^{-9} {\left(\frac{\alpha}{\pi} \right)^2}
 -6.5819(19) \times 10^{-8} {\left(\frac{\alpha}{\pi}
\right)^3} \nonumber
\end{eqnarray}
\vspace{-7mm}
\begin{eqnarray}
a_e(\mu,\tau)&=&0.190945(62) \times 10^{-12} {\left(\frac{\alpha}{\pi}
\right)^3} \;.
\nonumber
\end{eqnarray}
The QED part thus may be summarized in the prediction
\begin{eqnarray}
a_e^{\rm QED}
 &=& \frac{\alpha}{2\pi}
 -0.328\,478\,444\,002\,90(60) \left( \frac{\alpha}{\pi}\right)^2\nonumber\\
&&  \hspace*{-2cm}
 +1.181\, 234\, 016\, 828(19) \left( \frac{\alpha}{\pi}\right)^3
-1.9144(35)\left( \frac{\alpha}{\pi}\right)^4
+0.0(3.8)\left(\frac{\alpha}{\pi}\right)^5.
\label{02aeQED}
\end{eqnarray}
The hadronic and weak contributions to $a_e$ are small : $a_e^{\rm
had}=1.67(3) \times 10^{-12}$ and $a_e^{\rm weak}=0.036
\times 10^{-12}$, respectively. The hadronic contribution now just starts to be
significant, however, unlike in $a_\mu^{\rm had}$ for the muon,
$a_e^{\rm had}$ is known with sufficient accuracy and is not the
limiting factor here. The theory error is dominated by the missing
5-loop QED term.  As a result $a_e$ essentially only depends on perturbative
QED, while hadronic, weak and new physics (NP) contributions are suppressed by
$(m_e/M)^2$, where $M$ is a weak, hadronic or new physics scale. As a
consequence $a_e$ at this level of accuracy is theoretically well
under control (almost a pure QED object)
and therefore is an excellent observable for extracting
$\alpha_{\rm QED}$  based on the SM prediction
\begin{eqnarray}
a_e^{\rm SM} &=& a_e^{\rm QED} \mathrm{[Eq.~(\ref{02aeQED})]}
+1.706(30)\times 10^{-12}~~
\mbox{(hadronic \& weak)}\;.
\end{eqnarray}

We now compare this result with the very recent extraordinary precise measurement
of the electron anomalous magnetic moment\footnote{The famous $g_e$
measurement from University of Washington (Dehmelt et
al. 1987)~\cite{02aeold} found $a_{e}^{\rm exp} = 0.001\, 159\, 652 \, 188\, 30(420)$
and recently has been improved by about a factor 6 in an experiment at Harvard University
(Gabrielse et al. 2006). The new central value shifted downward by 1.7
standard deviations.}~\cite{02aenew}
\begin{equation}
a_{e}^{\rm exp}= 0.001\, 159\, 652\, 180\, 85(76)
\end{equation}
 which yields\\[-8mm]
\begin{eqnarray*}
\alpha^{-1}(a_e)&=&137.035999069(90)(12)(30)(3)\;,
\end{eqnarray*}
which is the value (\ref{02alphainv_a_e})~\cite{02alnew,02Aoyama07} we use in
calculating $a_\mu$.
The first error is the experimental one of $a_e^\mathrm{exp}$, the
second and third are the numerical uncertainties of the $\alpha^4$ and
$\alpha^5$ terms, respectively. The last one is the hadronic
uncertainty, which is completely negligible. Note that the largest
theoretical uncertainty comes from the almost completely missing
information concerning the 5--loop contribution. This is now the by
far most precise determination of $\alpha$ and we will use it
throughout in the calculation of $a_\mu$, below.\\
{\footnotesize The best
determinations of $\alpha$ which do not depend on $a_e$
are~\cite{02Cs06,02Rb06}
\begin{eqnarray*}
\alpha^{-1}(\mathrm{Cs06})&=&137.03600000(110)[8.0\, \mathrm{ppb}]\;,\\
\alpha^{-1}(\mathrm{Rb06})&=&137.03599884(091)[6.7\, \mathrm{ppb}]\;,
\label{02alphainv_ai}
\end{eqnarray*}
less precise by about a factor ten. $\alpha(\mathrm{Cs06})$ is determined from
a measurement of $h/M_\mathrm{Cs}$ via Cesium recoil measurements~\cite{02Cs06},
while $\alpha(\mathrm{Rb06})$ derives from the ratio
$h/M_\mathrm{Rb}$ measured via Bloch oscillations of Rubidium atoms
in an optical lattice~\cite{02Rb06}.
These values should be used in theoretical predictions of $a_e$.
Using $\alpha(\mathrm{Cs06})$ we get $a_e= 0.00115965217298(930)$ and
$a_e^\mathrm{exp}-a_e^\mathrm{the}=7.87(9.33) \times 10^{-12}$, with
$\alpha(\mathrm{Rb06})$ the prediction reads $a_e=
0.00115965218279(769)$ and
$a_e^\mathrm{exp}-a_e^\mathrm{the}=-1.94(7.73) \times 10^{-12}$ in
best agreement.  The error of the prediction is completely dominated
by the uncertainty coming from $\alpha(\mathrm{Cs06})$ and
$\alpha(\mathrm{Rb06})$ such that an improvement of $\alpha$ by a
factor 10 would allow a much more stringent test of QED (see also~\cite{02alnew,02Aoyama07}).
If one assumes that $\left| \Delta a_e^{\rm New\ Physics}\right|
\simeq m_e^2 / \Lambda^2$ where $\Lambda$ approximates the scale of
``New Physics'', then the agreement between $\alpha^{-1}(a_e)$ and
$\alpha^{-1}(\mathrm{Rb06})$ currently probes $\Lambda \lsim
O(\mbox{250 GeV})$. To access the much more interesting $\Lambda
\sim O(\mbox{1 TeV})$ region also a bigger effort on the theory
side would by necessary about the $O(\alpha^4)$ and the $O(\alpha^5)$ terms}.

\section{Standard Model Prediction for $a_\mu$}
\subsection{QED Contribution}

The SM prediction of $a_\mu$ looks formally very similar to the one
for $a_e$, however, besides the common universal part, the
mass dependent, the hadronic and the weak effects enter with very
different weight and significance. The mass-dependent QED corrections
follow from the universal set of diagrams (see
e.g. Fig.~\ref{02fig:threeloopdiag} for the 3 loop case) by replacing
the closed internal $\mu$--loops by $e$-- and/or $\tau$--loops. Typical
contributions come from vacuum polarization or light-by-light
scattering loops, like

\begin{figure}[h]
\vspace*{-3mm}
\centering
\includegraphics[height=2.2cm]{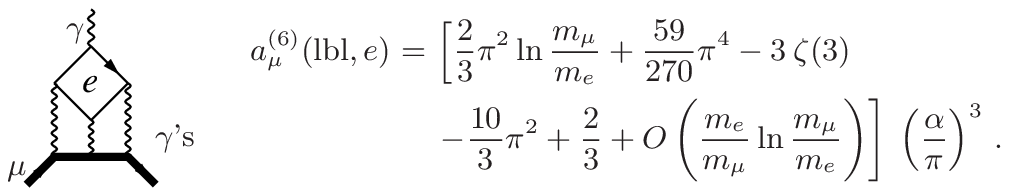}
\vspace*{-3mm}
\end{figure}
\noindent The result is given  by
\begin{equation}
a_\mu=a_e^{\rm uni}+a_\mu(m_\mu / m_e )+a_\mu(m_\mu /
m_\tau)+a_\mu(m_\mu /m_e,m_\mu / m_\tau)
\end{equation}
with\footnote{Again the order $\alpha^3$ terms are given by two parts
(see (\ref{02aeexpansion}))
\begin{eqnarray*}
A_2^{(6)}(m_\mu/m_e)
&=&\left. 20.947\,924\,89(16)\right|_{e - \mathrm{lbl}} +
\left. 1.920\,455\,130(33)\right|_{e - \mathrm{vap}} \\
A_2^{(6)}(m_\mu/m_\tau) &=&
\left. 0.002\,142\,83(69) \right|_{\tau - \mathrm{lbl}}
- \left. 0.001\,782\,33(48)\right|_{\tau - \mathrm{vap}}\; .
\end{eqnarray*}
The errors are due to the uncertainties in the mass ratios. Note that
the electron light-by-light scattering loop gives an unexpectedly
large contribution~\cite{02LBLlep69}.}~\cite{02El66,02La93,02LR93,02KOPV03}
\begin{eqnarray*}
a_\mu(m_\mu/m_e)&=&1.094~258~311~1~(84)\: { \left(\frac{\alpha}{\pi} \right)^2}
+ 22.868~380~02~(20)~  \:{ \left(\frac{\alpha}{\pi} \right)^3} \\&&
+~~ 132.682~3~(72)~~ \:
{ \left(\frac{\alpha}{\pi} \right)^4}
\\
a_\mu(m_\mu/m_\tau)&=&7.8064~(25) \times 10^{-5}~\: { \left(\frac{\alpha}{\pi} \right)^2}
+ 36.051~(21) \times 10^{-5}\: { \left(\frac{\alpha}{\pi} \right)^3} \\&&
+~~~~~ 0.005~(3)~~~\: {
\left(\frac{\alpha}{\pi} \right)^4}
\\
a_\mu(m_\mu/m_e,m_\mu/m_\tau)&=&
52.766~(17) \times 10^{-5}\: { \left(\frac{\alpha}{\pi} \right)^3}
+~~ 0.037~594~(83) \:
{ \left(\frac{\alpha}{\pi} \right)^4} \\
\end{eqnarray*}
except for the last term, which has been worked out as a series
expansion in the mass ratios~\cite{02CS99,02FGdR05}, all contributions
are known analytically in exact form~\cite{02La93,02LR93}\footnote{Explicitly, the papers only
present expansions in the mass ratios; some result have been extended in~\cite{02KOPV03}
and cross checked against the full analytic result in~\cite{02Passera04}.} up to
3--loops. At 4--loops only a few terms are known
analytically~\cite{02A8analytic}. Again the relevant 4--loop
contributions have been evaluated by numerical integration methods by
Kinoshita and Nio~\cite{02KiNi04}. The 5--loop term has been estimated
to be $A^{(10)}_{2}(m_\mu/m_e) = 663(2)$
in~\cite{02Ka93,02KiNi06,02Kataev05}.

Our knowledge of the QED result for $a_\mu$ may be summarized by
\begin{eqnarray}
a_\mu^{\rm QED}
 &=& \frac{\alpha}{2\pi}
 +0.765\,857\,410(26) \left( \frac{\alpha}{\pi}\right)^2\nonumber\\
&& \hspace*{-2cm}
 +24.050\,509\,65(46) \left( \frac{\alpha}{\pi}\right)^3
+130.8105(85)\left( \frac{\alpha}{\pi}\right)^4
+663(20)\left(\frac{\alpha}{\pi}\right)^5.
\end{eqnarray}
Growing coefficients in the $\alpha/\pi$ expansion reflect the
presence of large $\ln \frac{m_\mu}{m_e}\simeq 5.3$ terms coming from
electron loops. In spite of the strongly growing expansion coefficients
the convergence of the perturbation series is excellent

\begin{center}
\begin{tabular}{|c|r@{.}l|r|}
\hline
{ \# n of loops}& \multicolumn{2}{c|}{$C_i$ [$(\alpha/\pi)^n$]} & { $a_\mu^{\rm QED} \times 10^{11}~~~~~$} \\
\hline
1& +0&5 &  116140973.30 (0.08) \\
2& +0&765\,857\,410(26)&  413217.62 (0.01) \\
3& +24&050\,509\,65(46)&  30141.90 (0.00) \\
4& +130&8105(85)&  380.81 (0.03) \\
5& +663&0(20.0)&  4.48 (0.14)\\
\hline
{ tot}& \multicolumn{2}{c|}{~} & {  116584718.11 (0.16)}\\
\hline
\end{tabular}
 \end{center}
\noindent
because $\alpha/\pi$ is a truly small expansion parameter.

The different higher order
QED contributions are collected in Tab.~\ref{02tab:amuqedcontributions}.
We thus arrive at a QED prediction of $a_\mu$ given by
\begin{table}
\centering
\caption{QED contributions to $a_\mu$ in units $10^{-6}$}
\label{02tab:amuqedcontributions}
\begin{tabular}{l|r@{.}l|r@{.}l|r@{.}l|r@{.}l}
\hline\noalign{\smallskip}
\hline\noalign{\smallskip}
term & \multicolumn{2}{c}{universal} & \multicolumn{2}{|c}{$e$--loops}
& \multicolumn{2}{|c}{$\tau$--loops} & \multicolumn{2}{|c}{$e$\&$\tau$--loops} \\
\noalign{\smallskip}\hline\noalign{\smallskip}
$a^{(4)}$ & $\, -1$&$772\,305\,06~(0)    $&$\, 5$&$904\,060\,07~~(5)
          $&$\,  0$&$000\,421\,20(13)    $& \multicolumn{2}{c}{$-$} \\
$a^{(6)}$ & $    0$&$014\,804\,20~(0)    $&$   0$&$286\,603\,69~~(0)
          $&$    0$&$000\,004\,52~(1)    $&$\, 0$&$000\,006\,61(0)$ \\
$a^{(8)}$ & $   -0$&$000\,055\,73(10)    $&$   0$&$003\,862\,56~(21)
          $&$    0$&$000\,000\,15~(9)    $&$   0$&$000\,001\,09(0)$ \\
$a^{(10)} $&$    0$&$000\,000\,00(26)    $&$   0$&$000\,044\,83(135)
         $& \multicolumn{2}{c}{?}             & \multicolumn{2}{|c}{?} \\
\noalign{\smallskip}\hline
\end{tabular}
\end{table}
\begin{equation}
a_\mu^\mathrm{QED}=116\, 584\, 718.113(.082)(.014)(.025)(.137)[.162] \times 10^{-11}
\end{equation}
where the first error is the uncertainty of $\alpha$ in
(\ref{02alphainv_a_e}), the second one combines in quadrature the
uncertainties due to the errors in the mass ratios, the third is due
to the numerical uncertainty and the last stands
for the missing $O(\alpha^5)$ terms. With the new value of
$\alpha[a_e]$ the combined error is dominated by our limited knowledge of
the  5--loop term.

\subsection{Weak Contributions}
The electroweak SM is a non-Abelian gauge theory with gauge group
$SU(2)_L \otimes U(1)_Y \to U(1)_\mathrm{QED}$, which is broken down
to the electromagnetic Abelian subgroup $U(1)_\mathrm{QED}$ by the
Higgs mechanism, which requires a scalar Higgs field $H$ which
receives a vacuum expectation value $v$. The latter fixes the
experimentally well known Fermi constant $G_\mu=1/(\sqrt{2}v^2)$ and
induces the masses of the heavy gauge bosons $M_W$ and $M_Z$ as well
as all fermion masses $m_f$.  Other physical constants which we will
need later for evaluating the weak contributions are the Fermi
constant
\begin{equation}
G_\mu=1.16637(1) \times 10^{-5}~~~ \mathrm{GeV}^{-2}\; ,
\label{02gmu}
\end{equation}
the weak mixing parameter
\begin{equation}
\sin^2 \Theta_W=0.22276(56)
\label{02sin2W}
\end{equation}
and the masses of the intermediate gauge bosons $Z$ and $W$
\begin{equation}
M_Z=91.1876 \pm 0.0021 \ \mathrm{GeV} \; , M_W= 80.392 \pm 0.029 \ \mathrm{GeV} \; .
\label{02GBmasses}
\end{equation}
For the not yet discovered SM Higgs particle the mass is
constrained by LEP data to the range
\begin{equation}
114 \ \mathrm{GeV} < m_H <200 \ \mathrm{GeV \ } (\mathrm{at \ } 96\%
\mathrm{ \ CL}) \; .
\end{equation}
The weak interaction contributions to $a_\mu$ are due to the exchange
of the heavy gauge bosons, the charged $W^\pm$ and the neutral $Z$,
which mixes with the photon via a rotation by the weak mixing angle
$\Theta_W$ and which defines the weak mixing parameter $\sin^2
\Theta_W=1-M_W^2/M_Z^2$.  What is most interesting is the occurrence
of the first diagram of Fig.~\ref{02fig:oneloopweak}, which exhibits a
non-Abelian triple gauge vertex and the corresponding contribution
provides a test of the Yang--Mills
structure involved. It is of course not
surprising that the photon couples to the charged $W$ boson the way it
is dictated by electromagnetic gauge invariance.
\begin{figure}
\vspace*{-1mm}
\centering
\includegraphics[height=2cm]{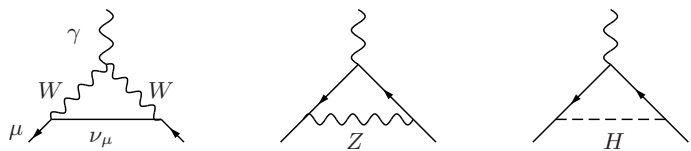}
\caption{The leading weak contributions to $a_\ell$; diagrams in the
physical unitary gauge}
\label{02fig:oneloopweak}
\end{figure}
The gauge boson contributions up to negligible terms of order
$O(\frac{m_\mu^2}{M_{W,Z}^2})$ are given by~\cite{02EW1Loop}
\begin{eqnarray}
a^{(2)\:\mathrm{EW}}_\mu(W)&=&\frac{\sqrt{2} G_\mu m_\mu^2}{16 \pi^2}\:\frac{10}{3}\simeq
+388.70(0) \times 10^{-11} \nonumber \\
a^{(2)\:\mathrm{EW}}_\mu(Z)&=&\frac{\sqrt{2} G_\mu m_\mu^2}{16 \pi^2}\:\frac{(-1+4\:\sin^2 \Theta_W)^2-5}{3}\simeq
-193.88(2) \times 10^{-11} \nonumber
\label{02oneloopbosonic}
\end{eqnarray}
while the diagram with the Higgs exchange, for $m_H \gg m_\mu$, yields
\begin{eqnarray*}
a^{(2)\:\mathrm{EW}}_\mu(H)
&\simeq& \frac{\sqrt{2} G_\mu m_\mu^2}{4 \pi^2} \:
\frac{m_\mu^2}{m_H^2}\:\ln \frac{m_\mu^2}{m_H^2 }+\cdots
\leq 5 \times 10^{-14}~~\mathrm{for}~~ m_H \geq  114~ \mathrm{GeV} \; .
\end{eqnarray*}
Employing the SM parameters given in (\ref{02gmu})
and (\ref{02sin2W}) we obtain
\begin{eqnarray}
a_\mu^{(2)\:\mathrm{EW}} = (194.82 \pm 0.02) \times 10^{-11}
\label{02EW1l}
\end{eqnarray}
The error comes from the uncertainty in $\sin^2\Theta_W$ given above.

The electroweak two--loop corrections have to be taken into account as
well. In fact triangle fermion--loops may give rise to unexpectedly
large radiative corrections. The diagrams which yield the leading
corrections are those including a VVA triangular fermion--loop
($VVA\neq0$ while $VVV=0$) associated with a $Z$ boson exchange
\\[-6mm]
\begin{center}
\includegraphics[height=2.3cm]{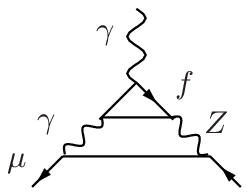}
\end{center}
which exhibits a parity violating axial coupling
(A). A fermion of flavor $f$ yields a
contribution
\begin{eqnarray}
a^{(4)\:\mathrm{EW}}_\mu([f])
\simeq \frac{\sqrt{2} G_\mu m_\mu^2}{16 \pi^2}
\frac{\alpha}{\pi}\:2T_{3f}N_{cf}Q_f^2\:\left[3 \ln \frac{M_Z^2}{m_{f'}^2} + C_f
\right]
\end{eqnarray}
where $T_{3f}$ is the 3rd component of the weak isospin, $Q_f$ the charge
and $N_{cf}$ the color factor, 1 for leptons, 3 for quarks. The mass
$m_{f'}$ is $m_\mu$ if $m_f < m_\mu$ and $m_f$ if $m_f > m_\mu$, and
$C_e=5/2,~C_\mu=11/6-8/9\:\pi^2,~C_\tau=-6$~\cite{02KKSS92}. However, in
the SM the consideration of individual fermions makes no sense and a
separation of quarks and leptons is not possible. Mathematical
consistency of the SM requires complete VVA anomaly cancellation
between leptons and quarks, and actually $\sum_fN_{cf}Q_f^2T_{3f}=0$
holds for each of the 3 known lepton--quark families separately.  Treating,
in a first step, the quarks like free fermions (quark parton model
QPM) the first two families yield (using $m_u=m_d=300~\mathrm{MeV} \;
, m_s=500~\mathrm{MeV} \; , m_c=1.5 \mathrm{GeV} $)
\begin{eqnarray}
a_\mu^{(4)\:\mathrm{EW}}(\left[\begin{tabular}{c}$e,u,d$ \\
$\mu,c,s$ \end{tabular}\right])_\mathrm{QPM} &\simeq&
-\frac{\sqrt{2} G_\mu\:m_\mu^2}{16 \pi^2}\:\frac{\alpha}{\pi}\:\left[
\ln \frac{m_u^8m_c^8}{m_\mu^{12}m_d^2m_s^2}+\frac{49}{3}-\frac{8\pi^2}{9}\right]
\nonumber \\ && \hspace*{-19mm} \simeq -\frac{\sqrt{2} G_\mu\:m_\mu^2}{16
\pi^2}\:\frac{\alpha}{\pi}\:\times 32.0(?) \simeq -8.65(?) \times 10^{-11} \;,
\end{eqnarray}
which demonstrates that the leading large logs $\sim \ln M_Z$ have
canceled~\cite{02CKM95F}, as it should be. However, the quark masses
which appear here are ill-defined constituent quark masses, which can
hardly account reliably for the strong interaction
effects, therefore the question marks in
place of the errors.

In fact, low energy QCD is characterized in the chiral limit of
massless light quarks $u,d,s$, by spontaneous chiral symmetry breaking
(S$\chi$SB) of the chiral group $SU(3)_V \otimes SU(3)_A$, which in
particular implies the existence of the
pseudoscalar octet of pions and
kaons as Goldstone bosons. The light quark
condensates are essential features in this situation and lead to
non-perturbative effects completely absent in a perturbative
approach. Thus low energy QCD effects are intrinsically
non--perturbative and controlled by chiral perturbation theory (CHPT),
the systematic QCD low energy expansion, which accounts for the
S$\chi$SB and the chiral symmetry breaking by quark masses in a
systematic manner. The low energy effective theory describing the
hadronic contributions related to the light quarks $u,d,s$ requires
the calculation of the diagrams of the type shown in
Fig.~\ref{02fig:XPTuds}.
\begin{figure}
\vspace*{-3mm}
\centering
\includegraphics[height=2.7cm]{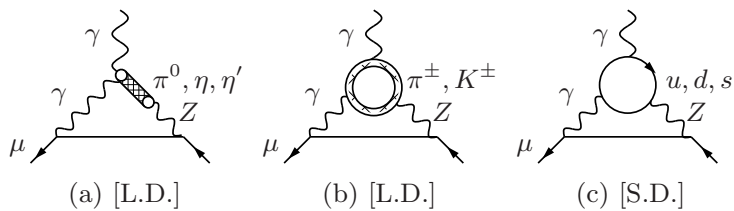}
\caption{The two leading CHPT diagrams (L.D.) and the
QPM diagram (S.D.). The charged pion loop is sub-leading and is
discarded. Diagrams with permuted $\gamma \leftrightarrow Z$ on the
$\mu$-line have to be included}
\label{02fig:XPTuds}
\end{figure}
The leading effect for the 1st plus 2nd family takes the form~\cite{02PPdeR95}
\begin{eqnarray}
a_\mu^{(4)\:\mathrm{EW}}(\left[\begin{tabular}{c}$e,u,d$ \\
$\mu,c,s$ \end{tabular}\right])_{\mathrm{CHPT}}&=&
\frac{\sqrt{2} G_\mu\:m_\mu^2}{16 \pi^2}\:\frac{\alpha}{\pi}\:\left[
-\frac{14}{3} \ln \frac{M_\Lambda^2}{m_\mu^2}+4\ln \frac{M_\Lambda^2}{m_c^2}
-\frac{35}{3}
+\frac{8}{9}\pi^2\right]
\nonumber \\ && \hspace*{-19mm} \simeq -\frac{\sqrt{2} G_\mu\:m_\mu^2}{16
\pi^2}\:\frac{\alpha}{\pi}\:\times 26.2(5) \simeq -7.09(13) \times 10^{-11} \; .
\label{02CHPTeudmsc}
\end{eqnarray}
The error comes from varying the cut--off $M_\Lambda$ between 1 GeV
and 2 GeV. Below 1 GeV CHPT can be trusted above 2 GeV we can trust
pQCD. Fortunately the result is not very sensitive to the choice of
the cut--off.  For more sophisticated analyses we refer
to~\cite{02CKM95F,02PPdeR95,02KPPdeR02} which was corrected and
refined in~\cite{02DG98,02CMV03}.
Thereby, a new kind of non-renormalization theorems played a key
role~\cite{02Vainshtein03,02KPPdR04,02JT05}. Including subleading
effects yields $-6.7\times 10^{-11}$ for the first two families. The
3rd family of fermions including the heavy top quark can be treated in
perturbation theory and has been worked out to be $-8.2\times
10^{-11}$ in~\cite{02D'Hoker92}. Subleading fermion loops contribute
$-5.3\times 10^{-11}$. There are many more diagrams contributing, in
particular the calculation of the bosonic contributions (1678
diagrams) is a formidable task and has been performed 1996 by
Czarnecki, Krause and Marciano as an expansion in $(m_\mu/M_V)^2$ and
$(M_V/m_H)^2$~\cite{02CKM96B}. Later complete calculations, valid also
for lighter Higgs masses, were performed~\cite{02HSW04,02GC05}, which
confirmed the previous result $-22.3\times 10^{-11}$. The 3rd family
of fermions including the heavy top quark can be treated in
perturbation theory and has been worked out
in~\cite{02D'Hoker92}.

The complete weak contribution may be summarized by~\cite{02CMV03}
\begin{eqnarray}
a_\mu^{\mathrm{EW}} &=&\frac{\sqrt{2} G_\mu\:m_\mu^2}{16
\pi^2}\:\left\{\frac53 + \frac13\:(1-4\sin^2 \Theta_W)^2-\frac{\alpha}{\pi}[155.5(4)(2)]\right\}
\nonumber \\ &=& (154 \pm 1 [\rm had] \pm 2 [m_H,m_t,3-loop])\times 10^{-11}
\label{02EWtot}
\end{eqnarray}
with errors from triangle quark loops and from  variation of the Higgs
mass in the range $m_H=150^{+100}_{-40}$ GeV. The 3-loop effect
has been estimated to be negligible~\cite{02DG98,02CMV03}.

\subsection{Hadronic Contributions}

So far when we were talking about fermion loops we only considered the
lepton loops. Besides the leptons also the strongly interacting quarks
have to be taken into account\footnote{The theory of strong
interactions is Quantum Chromodynamics (QCD)~\cite{02QCD}. The
strongly interacting particles, the hadrons, are made out of quarks
and/or antiquarks, which interact via an octet of gluons according to
the non-Abelian $SU(3)_c$ gauge theory. The gauged internal degrees of
freedom are named color. Quarks are flavored and labeled as up ($u$),
down ($d$), strange ($s$), charm ($c$), bottom ($b$) and top
($t$). Each of the flavored quarks exists in $N_c=3$ colors (red,
green, blue). All hadrons are color neutral bound states
(confinement).  This means that QCD is intrinsically
non-perturbative. However, QCD also has the property of asymptotic
freedom~\cite{02PGW73}, which implies that perturbation theory starts
to work at higher energies, where the quark structure appears resolved
as in deep inelastic electron-proton scattering, for example.}. The
problem is that strong interactions at low energy are non-perturbative
and straight forward first principle calculations become very
difficult and often impossible.\\[-5mm]

Fortunately the leading hadronic effects are vacuum polarization type
corrections (see (\ref{02runninge})), which can be safely evaluated by
exploiting causality (analyticity) and unitarity (optical theorem)
together with experimental low energy data. In fact vacuum
polarization effects may be calculated using the master formula
\begin{equation}
\frac{1}{q^2} \Rightarrow \int_0^\infty \frac{ds}{s}\frac{1}{q^2-s}
\:\frac{1}{\pi}\: \mbox{Im} \:\Pi_\gamma(s)
\label{02photondressing}
\end{equation}
which replaces a free photon propagator by a dressed one, and where
the imaginary part of the photon self-energy function $\Pi_\gamma(s)$
is determined via the optical theorem by the
total cross-section of hadron production in electron-positron
annihilation:
\begin{equation}
\sigma(s)_{e^+e^- \to \gamma^* \to \mathrm{hadrons}}=
\frac{4\pi^2\alpha}{s}\: \frac{1}{\pi}\: \mbox{Im} \:\Pi_\gamma(s)\;.
\label{02opticaltheorem}
\end{equation}
The leading hadronic contribution is represented by the diagram
Fig.~\ref{02fig:ammlohad},
\begin{figure}[h]
\vspace*{-3mm}
\centering
\includegraphics[height=2.7cm]{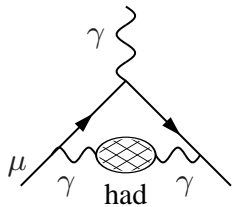}
\caption{The leading order (LO) hadronic vacuum polarization diagram}
\label{02fig:ammlohad}
\end{figure}

\noindent
which corresponds to a contribution $a_\mu^{\mathrm{massive \ }
\gamma}=K(s)$ of the lowest order diagram with the photon replaced by
a ``massive photon'' of mass $\sqrt{s}$, and convoluted according to
(\ref{02photondressing}).
It yields the dispersion integral
\begin{equation}
a_\mu=\frac{\alpha}{\pi}\:\int_0^\infty \frac{ds}{s}
\:\frac{1}{\pi}\: \mbox{Im} \:\Pi_\gamma(s)\:K(s)\;,\;\;
K(s)\equiv\int_0^1dx \frac{x^2(1-x)}{x^2+\frac{s}{m_\mu^2}(1-x)}\;.
\label{02amuhad2fold}
\end{equation}
As a result the leading non-perturbative hadronic contributions
$a_\mu^\mathrm{had}$ can be obtained in terms of
$R_\gamma(s) \equiv { \sigma^{(0)}(e^+e^- \rightarrow \gamma^*
\rightarrow {\rm hadrons})}/{ \frac{4\pi \alpha^2}{3s}}
$ data via the dispersion integral:\\[-4mm]
\begin{eqnarray}
a_\mu^\mathrm{had} &=& \left(\frac{\alpha m_\mu}{3\pi}
\right)^2 \bigg(\:
\int\limits_{4 m_\pi^2}^{E^2_{\rm cut}}ds\,
\frac{{ R^{\mathrm{data}}_\gamma(s)}\:\hat{K}(s)}{s^2}
+ \int\limits_{E^2_{\rm cut}}^{\infty}ds\,
\frac{{ R^{\mathrm{pQCD}}_\gamma(s)}\:\hat{K}(s)}{s^2}\,
\bigg)_{\,.}
\end{eqnarray}
The rescaled kernel function $\hat{K}(s)=3s/m_\mu^2\:K(s)$ is a smooth
bounded function, increasing from 0.63... at $s=4m_\pi^2$ to 1 as
$s\to \infty$. The $1/s^2$ enhancement at low energy implies that the
$\rho \to \pi^+\pi^-$ resonance is dominating the dispersion integral
($\sim$ 75 \%). Data can be used up to energies where $\gamma-Z$
mixing comes into play at about 40 GeV. However, by the virtue of
asymptotic freedom, perturbative Quantum Chromodynamics (pQCD) becomes
the more reliable the higher the energy and in fact may be used safely
in regions away from the flavor thresholds where the non-perturbative
resonances show up: $\rho$, $\omega$, $\phi$, the $J/\psi$ series and
the $\Upsilon$ series. We thus use perturbative QCD~\cite{02GKL,02ChK95}
from 5.2 to 9.46 GeV and for the high energy tail above 13 GeV, as
recommended in~\cite{02GKL,02ChK95,02HS02}.

Hadronic cross section measurements $e^+e^- \to \mathrm{hadrons}$ at
electron-positron storage rings started in the early 1960's and
continued up to date. Since our analysis~\cite{02EJ95} in 1995 data
from MD1~\cite{02MD196}, BES-II~\cite{02BES} and from
CMD-2~\cite{02CMD2} have lead to a substantial reduction in the
hadronic uncertainties on $a_\mu^\mathrm{had}$. More recently,
KLOE~\cite{02Aloisio:2004bu}, SND~\cite{02Achasov:2006vp} and
CMD-2~\cite{02Aulchenko:2006na} published new measurements in the
region below 1.4 GeV. My up-to-date
evaluation of the leading order hadronic VP yields~\cite{02FJ06}
\begin{equation}
a_\mu^\mathrm{had(1)}=(692.1\pm 5.6)\:\times 10^{-10}\; .
\end{equation}
Some other recent evaluations are collected in Tab.~\ref{02tab:otheramuhad1}.
Differences in errors
\begin{table}
\caption{Some recent evaluations of $a_\mu^{\rm had (1)}$}
\label{02tab:otheramuhad1}
\vspace*{-0mm}
\centering
\parbox{1.2in}{%
\hspace*{-2.5cm}%
\begin{tabular}{llc}
\hline\noalign{\smallskip}
 $a_\mu^{\rm had (1)}\times 10^{10}$ & data & Ref. \\
\noalign{\smallskip}\hline\noalign{\smallskip}
$ 696.3[7.2]$ &$e^+e^-$&\cite{02DEHZ03}\\
$ 711.0[5.8]$ &$e^+e^-+\tau$  &\cite{02DEHZ03}\\
$ 694.8[8.6]$ &$e^+e^-$&\cite{02GJ03} \\
$ 684.6[6.4]$ & $e^+e^-$ TH&\cite{02SN03} \\
$ 699.6[8.9]$ &$e^+e^-$&\cite{02ELZ03}\\
$ 692.4[6.4]$ &$e^+e^-$&\cite{02HMNT04}\\
$ 693.5[5.9]$ &$e^+e^-$&\cite{02TY04}\\
$ 701.8[5.8]$ &$e^+e^-+\tau$&\cite{02TY04}\\
$ 690.9[4.4]$ &$e^+e^-$$^{**}$&\cite{02DEHZ06}\\
$ 689.4[4.6]$ &$e^+e^-$$^{**}$&\cite{02HMNT06}\\
$ 692.1[5.6]$ &$e^+e^-$$^{**}$&\cite{02FJ06} \\
\noalign{\smallskip}\hline
\end{tabular}}
\qquad
\begin{minipage}{1.2in}%
\vspace*{8.4mm}
\hspace*{-1.9cm}%
\includegraphics[height=5.0cm]{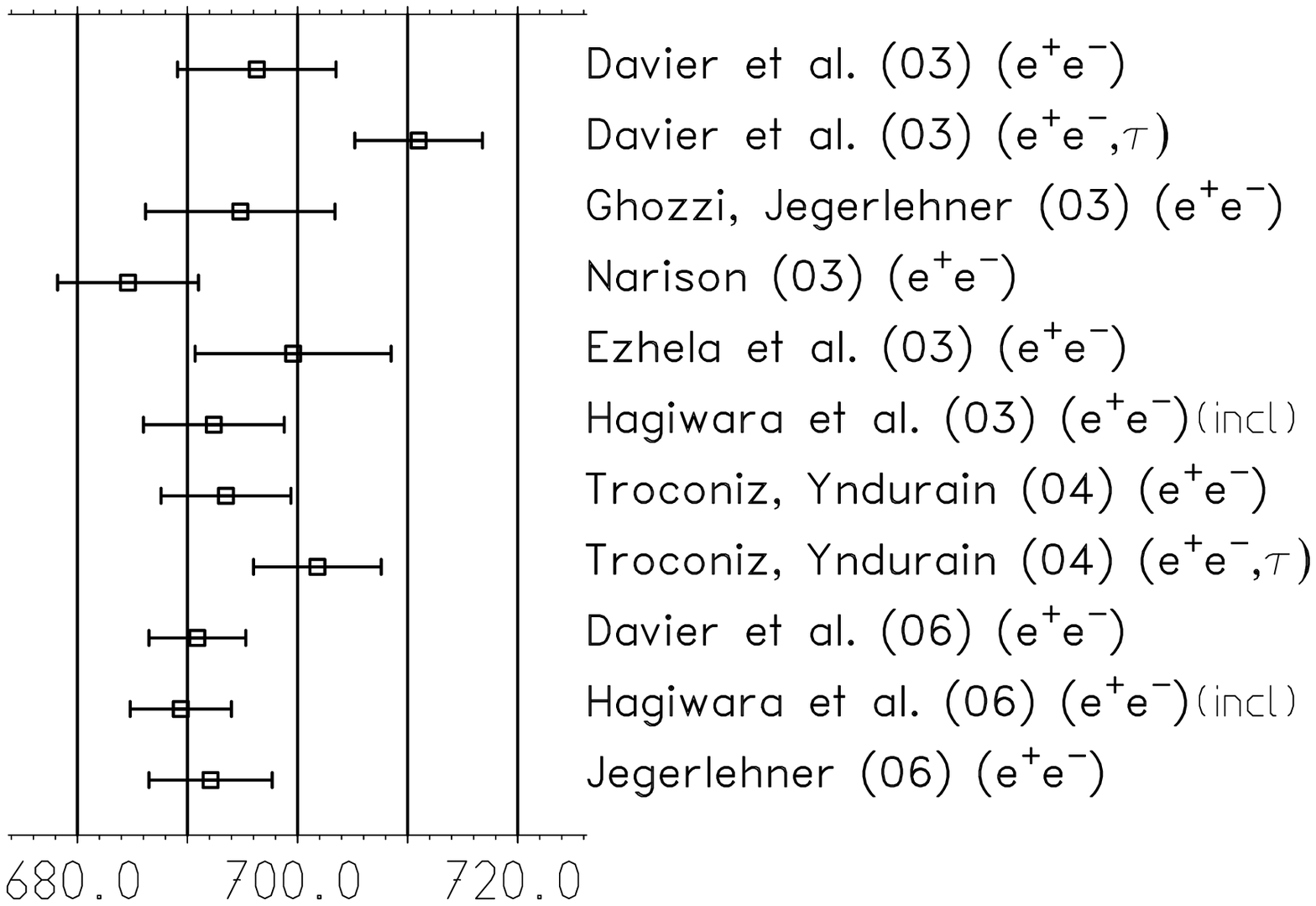}
\end{minipage}%
\vspace*{-2mm}
\end{table}

\noindent come about mainly by utilizing more
``theory-driven'' concepts : use of selected data sets only, extended
use of perturbative QCD in place of data [assuming local duality], sum
rule methods and low energy effective methods~\cite{LeCo02}. Only the
last three ($^{**}$) results include the most recent data from SND,
CMD-2, and BaBar\footnote{The analysis~\cite{02HMNT06} does not
include exclusive data in a range from 1.43 to 2 GeV; therefore also
the new BaBar data are not included in that range. It also should be
noted that CMD-2 and SND are not fully independent measurements; data
are taken at the same machine and with the same radiative correction
program. The radiative corrections play a crucial role at the present
level of accuracy, and common errors have to be added
linearly. In~\cite{02DEHZ03,02DEHZ06} pQCD is used in the extended
ranges 1.8 - 3.7 GeV and above 5.0 GeV; furthermore~\cite{02DEHZ06}
excludes the KLOE data.}.

In principle, the $I=1$ iso-vector part of $e^+e^- \to
\mathrm{hadrons}$ can be obtained in an alternative way by using the
precise vector spectral functions from hadronic $\tau$--decays $\tau
\to \nu_\tau + \mathrm{hadrons}$ which are related by an isospin
rotation~\cite{02ADH98}. After isospin violating corrections, due to
photon radiation and the mass splitting $m_d-m_u \neq 0$, have been
applied, there remains an unexpectedly large discrepancy between the
$e^+e^-$- and the $\tau$-based determinations of
$a_\mu$~\cite{02DEHZ03}, as may be seen in
Table~\ref{02tab:otheramuhad1}. Possible explanations are so far
unaccounted isospin breaking~\cite{02GJ03} or experimental problems
with the data. Since the $e^+e^-$-data are more directly related to
what is required in the dispersion integral, one usually advocates to
use the $e^+e^-$ data only.

At order $O(\alpha^3)$ diagrams of the type shown in
Fig.~\ref{02fig:ammhohad} have to be calculated, where the first
diagram stands for a class of higher order hadronic contributions
obtained if one replaces in any of the first 6 two--loop diagrams on
p.~\pageref{02twoloopuni} one internal photon line by a dressed one.
\begin{figure}
\vspace*{-3mm}
\centering
\includegraphics[height=2cm]{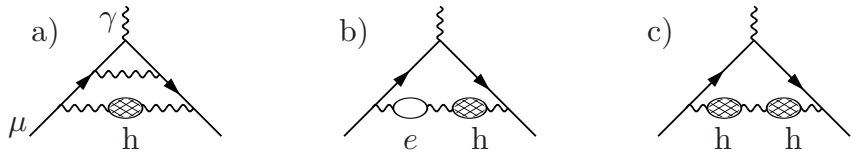}
\caption{Higher order (HO) vacuum polarization contributions}
\label{02fig:ammhohad}
\end{figure}
The relevant kernels for the corresponding dispersion integrals have
been calculated analytically in~\cite{02BR74} and appropriate series
expansions were given in~\cite{02Krause96} (for earlier estimates see
~\cite{02Calmet76,02KNO84}). Based on my recent compilation of the
$e^+e^-$ data~\cite{02FJ06} I obtain
\begin{equation}
a_\mu^\mathrm{had(2)}=(-100.3\pm 2.2)\:\times 10^{-11}\; ,
\end{equation}
in accord with previous/other
evaluations~\cite{02KNO84,02Krause96,02ADH98,02HMNT04,02HMNT06}.

Much more serious problems with non-perturbative hadronic effect we
encounter with the hadronic light-by-light (LbL) contribution at
$O(\alpha^3)$ depicted in Fig.~\ref{02fig:amulblhad}.
\begin{figure}
\vspace*{-3mm}
\centering
\includegraphics[height=2.7cm]{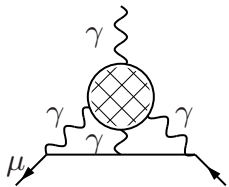}
\caption{Hadronic light-by-light scattering in $g-2$}
\label{02fig:amulblhad}
\end{figure}
Experimentally, we know that $\gamma \gamma \to \mathrm{hadrons} \to
\gamma \gamma$ is dominated by the hadrons
$\pi^0,~\eta,~\eta',\cdots$, i.e., single pseudoscalar meson
spikes~\cite{02LBLfacts}, and that $\pi^0 \to \gamma \gamma$ etc. is
governed by the parity odd Wess-Zumino-Witten (WZW) effective
Lagrangian
\begin{equation}
{\cal L}^{(4)}=-\frac{\alpha N_c}{12\ \pi f_0}\varepsilon  _{\mu \nu \rho \sigma}
F^{\mu \nu} A^\rho \partial^\sigma \pi^0 + \cdots
\end{equation}
which reproduces the Adler-Bell-Jackiw triangle
anomaly and which helps in estimating the
leading hadronic LbL contribution. $f_0$ denotes the pion decay
constant $f_\pi$ in the chiral limit of massless light quarks. Again,
in a low energy effective description, the quasi Goldstone
bosons, the pions and kaons play
an important role, and the relevant diagrams are displayed in
Fig~\ref{02fig:LbLpionpole}.
\begin{figure}
\vspace*{-0mm}
\centering
\includegraphics[height=2.7cm]{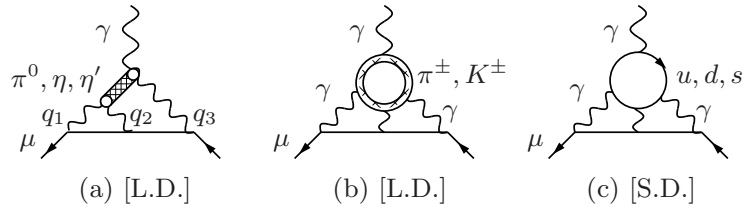}
\caption{Leading hadronic light--by--light scattering diagrams:
the two leading CHPT diagrams (L.D.) and the
QPM diagram (S.D.). The charged pion loop is sub-leading only, actually.
Diagrams with permuted $\gamma$'s on the
$\mu$-line have to be included. $\gamma$-hadron/quark vertices at
$q^2\neq0$ are dressed (VMD)}
\label{02fig:LbLpionpole}
\end{figure}

However, as we know from the hadronic VP discussion, the $\rho$ meson
is expected to play an important role in the game.  It looks natural
to apply a vector-meson dominance (VMD) like model.  Electromagnetic
interactions of pions treated as point-particles would be descried by
scalar QED in a first step.  However, due to hadronic interactions the
photon mixes with hadronic vector-mesons like the $\rho^0$. The naive
VMD model attempts to take into account this hadronic dressing by
replacing the photon propagator as $$\frac{\I\: g^{\mu \nu}}{q^2} +
\cdots \to
\frac{\I\: g^{\mu \nu}}{q^2} +\cdots -\frac{\I\:(g^{\mu
\nu}-\frac{q^\mu q^\nu}{m_\rho^2})}{q^2-m_\rho^2} =\frac{\I\: g^{\mu
\nu}}{q^2}\:\frac{m_\rho^2}{q^2-m_\rho^2} + \cdots \,,$$ where the
ellipses stand for the gauge terms. The main effect is that
it provides a damping at high energies with the $\rho$ mass as an
effective cut-off (physical version of a Pauli-Villars cut-off).
However, the naive VMD model is not compatible with chiral
symmetry. The way out is the Resonance Lagrangian Approach
(RLA)~\cite{02EGPdeR89} , an extended version of CHPT which incorporates
vector-mesons in accordance with the basic symmetries. The Hidden Local
Symmetry (HLS)~\cite{02HLS88} model and the Extended Nambu-Jona-Lasinio (ENJL)
~\cite{02ENJL96} model are alternative versions of RLA, which are basically
equivalent~\cite{02Prades99}, for what concerns this application.

Based on such effective field theory (EFT) models, two major efforts
in evaluating the full $a_\mu^{\mathrm{LbL}}$ contribution were made
by Hayakawa, Kinoshita and Sanda (HKS 1995)~\cite{02HKS95}, Bijnens,
Pallante and Prades (BPP 1995)~\cite{02BijnensLBL} and Hayakawa and
Kinoshita (HK 1998)~\cite{02HK98} (see also Kinoshita, Nizic and Okamoto
(KNO 1985)~\cite{02KNO84}). Although the
details of the calculations are quite different, which results in a
different splitting of various contributions, the results are in good
agreement and essentially given by the $\pi^0$-pole contribution,
which was taken with the wrong sign, however. In order to eliminate
the cut-off dependence in separating L.D. and S.D. physics, more
recently it became favorable to use quark--hadron duality, as it holds
in the large $N_c$ limit of QCD, for modeling of the hadronic
amplitudes~\cite{02deRafaelEJLN94}. The infinite series of narrow vector
states known to show up in the large $N_c$ limit is then approximated
by a suitable lowest meson dominance (LMD+V) ansatz~\cite{02LMD98},
assumed to be saturated by known low lying physical states of
appropriate quantum numbers. This approach was adopted in a reanalysis
by Knecht and Nyffeler (KN 2001)~\cite{02KnechtNyffeler01,02KNPdeR01,02BCM02,02RMW02} in
2001, in which they discovered a sign mistake in the dominant
$\pi^0,\eta,\eta'$ exchange contribution, which changed the central
value by $+167 \times 10^{-11}$, a 2.8 $\sigma$ shift, and which reduces a
larger discrepancy between theory and experiment.  More recently
Melnikov and Vainshtein (MV 2004)~\cite{02MV03} found additional
problems in previous calculations, this time in the short distance
constraints (QCD/OPE) used in matching the high energy behavior of the
effective models used for the $\pi^0,\eta,\eta'$ exchange
contribution.

The\footnote{This paragraph cannot be more than a rough sketch
of an ongoing discussion.}  most important pion-pole term
is of the form ($p$ is the muon
momentum, $q_i$ ($i=1,2,3$) are the virtual photon momenta, two of
which are chosen as loop integration variables)~\cite{02KnechtNyffeler01}
\begin{eqnarray}
a_{\mu}^{\mbox{\tiny{LbL;$\pi^0$}}}& = & - e^6
\int \frac{\D^4 q_1}{(2\pi)^4} \frac{\D^4 q_2}{(2\pi)^4}
\,\frac{1}{q_1^2 q_2^2 (q_1 + q_2)^2[(p+ q_1)^2 - m^2][(p - q_2)^2 - m^2]}
\nonumber \\
&& \quad \quad \times \left[
\frac{{\cal F}_{\pi^* \gamma^* \gamma^*}(q_2^2, q_1^2, q_3^2) \
{\cal F}_{\pi^* \gamma^* \gamma}(q_2^2, q_2^2, 0)}{q_2^2 -
m_{\pi}^2+\I \varepsilon} \ T_1(q_1,q_2;p) \nonumber \right. \\
&& \quad \quad \quad + \left.
\frac{{\cal F}_{\pi^* \gamma^* \gamma^*}(q_3^2, q_1^2,  q_2^2) \
{\cal F}_{\pi^* \gamma^* \gamma}(q_3^2, q_3^2,
0)}{q_3^2 - m_{\pi}^2+\I \varepsilon} \ T_2(q_1,q_2;p) \right] \,,
\label{02amulblpi0int}
\end{eqnarray}
where $T_1(q_1,q_2;p)$ and $T_2(q_1,q_2;p)$ are known scalar
kinematics factors and ${\cal F}_{\pi^* \gamma^* \gamma^*}(q_1^2,
q_2^2, q_3^2)$ is the non-perturbative $\pi^0 \gamma \gamma$ form
factor (FF) whose off-shell form is essentially unknown in the
integration range of (\ref{02amulblpi0int}).

A new quality of the problem encountered here is the fact that the
integrand depends on 3 invariants $q_1^2$, $q_2^2$, $q_3^2$
$q_3=-(q_1+q_2)$.  While hadronic VP correlators or the VVA triangle
with an external zero momentum vertex only depends on a single
invariant $q^2$. In the latter case the invariant amplitudes (form
factors) may be separated into a low energy part $q^2\leq \Lambda^2$
(soft) where the low energy effective description applies and a high
energy part $q^2 > \Lambda^2$ (hard) where pQCD works. In multi-scale
problems, however, there are mixed soft-hard regions where no answer
is available in general, unless we have data to constrain the
amplitudes in such regions. In our case, only the soft region
$q_1^2,q_2^2,q_3^2 \leq \Lambda^2$ and the hard region
$q_1^2,q_2^2,q_3^2 > \Lambda^2$ are under control of either the low
energy EFT and of pQCD, respectively. In the mixed soft-hard domains
operator product expansions and/or soft versus hard factorization
``theorems'' \`a la Brodsky-Farrar~\cite{02BrodskyFarrar73} may help.
Actually, one more approximation is usually made: the {\em pion-pole
approximation} ,i.e., the pion-momentum square (first argument of
${\cal F}$) is set equal to $m_\pi^2$, as the main contribution is
expected to come from the pole. Knecht and Nyffeler modeled ${\cal
F}_{\pi \gamma^*
\gamma^*}(m_\pi^2, q_1^2, q_2^2)$ in the spirit of the large $N_c$
expansion as a ``LMD+V'' form factor:\\
\begin{equation} \scriptsize
{\cal F}_{\pi \gamma^* \gamma^*}(m_\pi^2,q_1^2,q_2^2)
= \frac{ f_\pi}{3}~
\frac{q_1^2q_2^2(q_1^2+q_2^2)+h_1(q_1^2+q_2^2)^2
+h_2 q_1^2 q_2^2 + h_5 (q_1^2+q_2^2)+h_7
}{(q_1^2 -M_1^2)(q_1^2-M_2^2)(q_2^2-M_1^2)(q_2^2-M_2^2)}\,,
\label{02lmdv}
\end{equation}
with $h_7=-(N_cM_1^4M_2^4/4\pi^2 f_\pi^2)$, $f_\pi
 \simeq92.4~\mathrm{MeV}$.  An important constraint comes from the
pion-pole form factor ${\cal F}_{\pi \gamma^* \gamma}(m_\pi^2,
-Q^2,0)$, which has been measured by CELLO~\cite{02CELLO90} and
CLEO~\cite{02CLEO98}. Experiments are in fair agreement with the
Brodsky--Lepage~\cite{02LepageBrodsky80} form
\begin{equation}
{\cal F}_{\pi \gamma^* \gamma}(m_\pi^2, -Q^2,
0) \simeq - \frac{N_c}{12 \pi^2 f_\pi} \frac{1}{1+(Q^2/8\pi^2 f_\pi^2)}
\label{02FFCC}
\end{equation}
which interpolates between a $1/Q^2$ asymptotic behavior and the
constraint from $\pi^0$ decay at $Q^2=0$. This behavior requires
$h_1=0$. Identifying the resonances with $M_1=M_\rho =769~{\rm MeV}$,
$M_2=M_{\rho'} = 1465~{\rm MeV}$, the phenomenological constraint
fixes $h_5 =6.93~{\rm GeV}^4$. $h_2$ will be fixed by later.  As the
previous analyses, Knecht and Nyffeler apply the above VMD type form
factor on both ends of the pion line. In fact at the vertex attached
to the external zero momentum photon, this type of pion-pole form
factor cannot apply for kinematical reasons: when
$q^\mu_\mathrm{ext}=0$ not ${\cal F}_{\pi \gamma^*
\gamma}(m_\pi^2, -Q^2, 0)$ but ${\cal F}_{\pi^* \gamma^*
\gamma}(q_2^2, q_2^2,0)$ is the relevant object to be used, where
$q_2$ is to be integrated over. However, for large $q_2^2$ the pion
must be far off-shell, in which case the pion exchange effective
representation becomes obsolete. Melnikov and Vainshtein reanalyzed
the problem by performing an operator product expansion (OPE) for
$q_1^2 \simeq q_2^2 \gg (q_1+q_2)^2 \sim m_\pi^2$. In the chiral limit
this analysis reveals that the external vertex is determined by the
exactly known ABJ anomaly $ {\cal F}_{\pi
\gamma \gamma}(m_\pi^2, 0, 0)=-1/(4\pi^2 f_\pi)$. This means that
in the chiral limit there is no VMD like damping at high energies at the
external vertex. However, the absence of a damping in the chiral limit
does not prove that there is no damping in the real world with
non-vanishing quark masses. In fact, the quark triangle-loop
in this case provides a representation of the $\pi^{0*}
\gamma^* \gamma^*$ amplitude given by
\begin{eqnarray}
F^\mathrm{CQM}_{\pi^{0*} \gamma^* \gamma^*}(q^2,p_1^2,p_2^2)&\equiv&
(-4\pi^2 f_\pi)\: {\cal F}_{\pi^*
\gamma^* \gamma^*}(q^2,p_1^2,p_2^2)=
2m_q^2\:C_0(m_q;q^2,p_1^2,p_2^2)\nonumber \\ &\equiv& \int[\D
\alpha]\:\frac{2m_q^2} {m_q^2-\alpha_2 \alpha_3 p_1^2-\alpha_3
\alpha_1 p_2^2-\alpha_1 \alpha_2 q^2}\; ,
\end{eqnarray}
where $[\D \alpha] = \D \alpha_1\D \alpha_2\D
\alpha_3\:\delta(1-\alpha_1-\alpha_2-\alpha_3)$ and $m_q$ is a
constituent quark mass ($q=u,d,s$). For $p_1^2=p_2^2=q^2=0$ we obtain
$F^\mathrm{CQM}_{\pi^{0*} \gamma^* \gamma^*}(0,0,0)=1$, which is the
proper ABJ anomaly. Note the
symmetry of $C_0$ under permutations of the arguments ($p_1^2,p_2^2,q^2$).
For large $p_1^2$ at $p_2^2 \sim 0,~q^2 \sim 0$ or
$p_1^2 \sim p_2^2$ at $q^2 \sim 0$ the asymptotic
behavior is given by
\begin{eqnarray}
F^\mathrm{CQM}_{\pi^{0} \gamma^* \gamma}(0,p_1^2,0)\sim
r\:
\ln^2 r\;,\;\;
F^\mathrm{CQM}_{\pi^{0} \gamma^* \gamma^*}(0,p_1^2,p_1^2)\sim
2\:r\:
\ln r
\end{eqnarray}
where $r=\frac{m_q^2}{-p_1^2}$. The same behavior follows for $q^2
\sim p_1^2$ at $p_2^2 \sim 0$.  Note that in all cases we have the
same power behavior $\sim m_q^2/p_i^2$ modulo logarithms. Thus at high
energies the anomaly gets screened by chiral symmetry breaking
effects.

We therefore advocate to use consistently dressed form factors as
inferred from the resonance Lagrangian approach.
However, other effects which were first considered in~\cite{02MV03} must
be taken into account:\\
{\bf 1)} the constraint on the twist four $(1/q^4)$-term in the OPE requires
$h_2=-10$ GeV$^2$ in the Knecht-Nyffeler from factor (\ref{02lmdv}):
$\delta a_\mu\simeq +5\pm0$\\
{\bf 2)} the contributions from the $f_1$ and $f_1'$ isoscalar axial-vector mesons:
$\delta a_\mu\simeq +10\pm4$ (using dressed photons)\\
{\bf 3)} for the remaining effects: scalars ($f_0$) $+$ dressed $\pi^\pm,K^\pm$ loops $+$
dressed quark loops: $\delta a_\mu\simeq +2\pm6$\\
Note that the remaining terms have been evaluated in~\cite{02HKS95,02BijnensLBL} only.
The splitting into the different terms is model dependent and only the sum
should be considered: the results read $-5\pm13$ (BPP) and $5.2\pm13.7$ (HKS)
and hence the true contribution remains unclear\footnote{The problem seems
to be the sizable negative scalar contribution of~\cite{02BijnensLBL}, which
in~\cite{02KNO84} was estimated to be much smaller. Also the sign seems
to be in question.}.\\

An overview of results is presented in
Table~\ref{02tab:LbLrecent}. The last column gives my estimates base
on~\cite{02HKS95,02BijnensLBL,02KnechtNyffeler01,02MV03}. The ``no
FF'' column shows results for undressed photons (no form factor).
\begin{table}[t]
\centering
\caption{LbL: Summary of most recent results for $a_\mu \times 10^{11}$}
\label{02tab:LbLrecent}
\begin{tabular}{c|r|r|r|r|r|r}
\hline\noalign{\smallskip}
 &\multicolumn{1}{c|}{\rm no FF}& \multicolumn{1}{c|}{\rm BPP} & \multicolumn{1}{c|}{\rm HKS} & \multicolumn{1}{c|}{\rm KN} & \multicolumn{1}{c|}{\rm MV} &\multicolumn{1}{c}{\rm FJ} \\
\noalign{\smallskip}\hline\noalign{\smallskip}
 $\pi^0,\eta,\eta'$ & $+\infty~~$ & $85 \pm 13~$ & $82.7 \pm 6.4$ & $83 \pm 12$ & $114 \pm 10$& $88 \pm 12$ \\
axial vector &  & $2.5 \pm 1.0$ & $1.7 \pm 0.0$ & & $22 \pm ~5$& $10 \pm ~4$ \\
scalar       &  & $-6.8 \pm 2.0$ & \multicolumn{1}{c|}{$-$} &
\multicolumn{1}{c|}{$-$} &\multicolumn{1}{c|}{$-$} & $0\pm ~7$ \\
$\pi,K$ loops &$-49.8~$ &$-19 \pm 13~$ & $-4.5 \pm 8.1$ & & $0\pm 10$& $-19\pm 13$ \\
quark loops  &$~~62(3)~$& $21 \pm ~3~$ & $9.7 \pm 11.1$ & \multicolumn{1}{c|}{$-$} & \multicolumn{1}{c|}{$-$} & $21\pm ~3$ \\
\noalign{\smallskip}\hline
total & & $83\pm 32~$ & $89.6 \pm 15.4$ & $80 \pm 40$ & $136 \pm 25$& $100 \pm 39$ \\
\noalign{\smallskip}\hline
\end{tabular}
\end{table}
The constant WZW
form factor yields a divergent result, applying a cut-off $\Lambda$
one obtains~\cite{02KNPdeR01}
$(\alpha/\pi)^3{\cal C} \ln^2 \Lambda$, with an universal coefficient
${\cal C}=N_c^2m_\mu^2/(48 \pi^2 f_\pi^2)$; in the VMD dressed cases $M_V$
represents the cut-off $\Lambda \to M_V$ if $M_V \to \infty$.

\section{Theory Confronting the Experiment}

The following Tab.~\ref{02tab:amcontributions} collects the
typical contributions to $a_\mu$ evaluated in terms of $\alpha$
determined via  $a_e$ (\ref{02alphainv_a_e}).
\begin{table}
\caption{The various types  of contributions to $a_\mu$ in units $10^{-6}$,
ordered according to their size (L.O. lowest order, H.O. higher order,
LbL. light--by--light)}
\label{02tab:amcontributions}
\hspace*{5mm}
\vspace*{-0mm}
\centering
\parbox{1.2in}{%
\hspace*{-2.5cm}%
\begin{tabular}{l|r@{.}l}
\hline\noalign{\smallskip}
\hline\noalign{\smallskip}
  L.O. universal   & $~~1161$&$409\,73~~~(0) $\\
 $e$--loops        & $     6$&$194\,57~~~(0) $\\
 H.O. universal    & $    -1$&$757\,55~~~(0) $\\
 L.O. hadronic     & $     0$&$069\,21~~(56) $\\
 L.O. weak         & $     0$&$001\,95~~~(0) $\\
 H.O. hadronic     & $    -0$&$001\,00~~~(2) $\\
 LbL. hadronic     & $     0$&$001\,00~~(39) $\\
$\tau$--loops      & $     0$&$000\,43~~~(0) $\\
 H.O. weak         & $    -0$&$000\,41~~~(2) $\\
$e$+$\tau$--loops  & $     0$&$000\,01~~~(0) $\\
\noalign{\smallskip}\hline\noalign{\smallskip}
 theory     & $1165$&$917\,93~~(68)$ \\
 experiment & $1165$&$920\,80~~(63)$ \\
\noalign{\smallskip}\hline
\end{tabular}}
\qquad
\begin{minipage}{1.2in}%
\vspace*{1.8mm}
\hspace*{-1.0cm}%
\includegraphics[width=6.1cm,height=5.2cm]{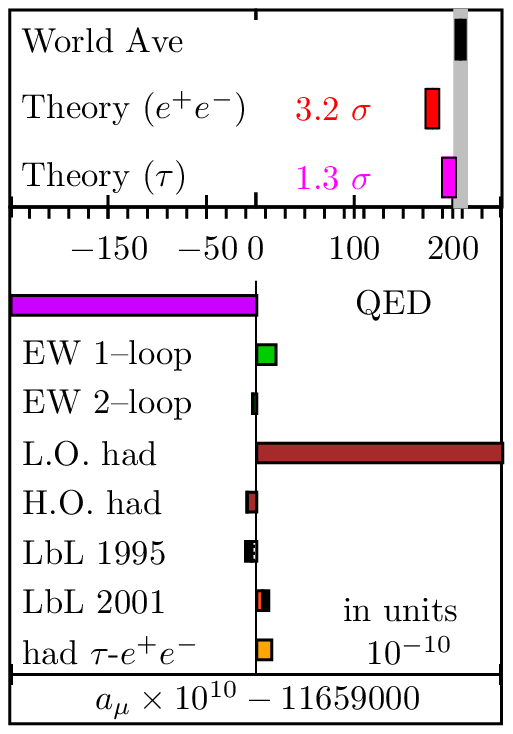}
\end{minipage}%
\vspace*{-2mm}
\end{table}
The world average experimental muon magnetic anomaly, dominated by the
very precise BNL result, now is~\cite{02BNL04}
\begin{equation}
a_\mu^\mathrm{exp} =1.165 920 80(63) \times
10^{-3}
\label{02finalamu}
\end{equation}
(relative uncertainty $5.4 \times
10^{-7}$), which confronts the SM prediction
\begin{equation}
a_\mu^\mathrm{the} =1.165 917 93(68)
\times 10^{-3} \; .
\end{equation}
Fig.~\ref{02fig:14} illustrates the improvement achieved by
the BNL experiment. The theoretical predictions mainly differ by the
L.O. hadronic effects, which also dominate the theoretical error.
A deviation between theory and experiment of about 3 $\sigma$ was persisting
since the first precise BNL result was released in 2000, in spite of
progress in theory and experiment since.

\begin{figure}[t]
\centering
\includegraphics[height=9cm]{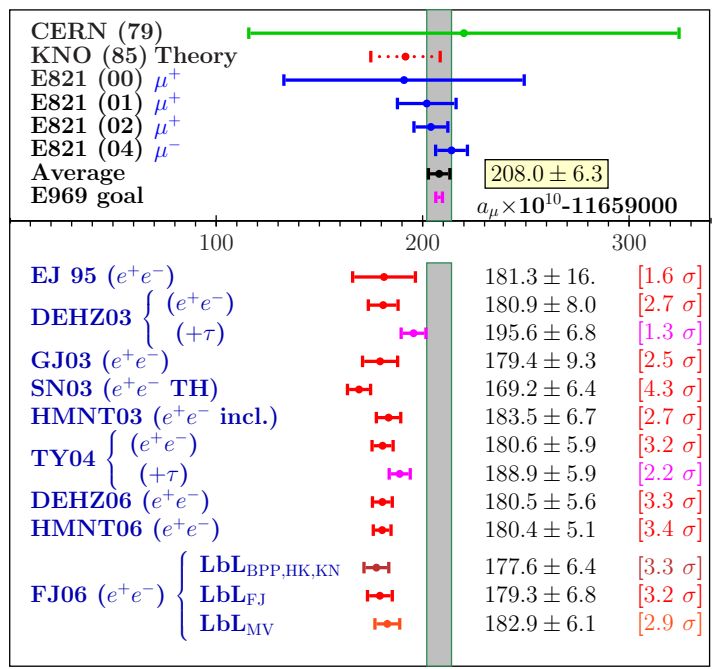}
\caption{Comparison between theory and experiment. Results differ by
different L.O. hadronic vacuum polarizations and variants of the LbL
contribution. Some estimates include isospin rotated $\tau$--data
($+\tau$)). The last entry FJ06 also illustrates the effect of using
different LbL estimations: 1) Bijnens, Pallante, Prades
(BPP)~\cite{02BijnensLBL}, Hayakawa, Kinoshita (HK)~\cite{02HK98} and Knecht,
Nyffeler (KN)~\cite{02KnechtNyffeler01}; 2) my estimation based on the other
evaluations; 3) the Melnikov, Vainshtein (MV)~\cite{02MV03} estimate of
the LbL contribution. EJ95 vs. FJ06 illustrates the improvement of the
$e^+e^-$-data between 1995 and now (see also
Tab.~\ref{02tab:otheramuhad1}). E969 is a
possible follow-up experiment of E821 proposed
recently~\cite{02LeeRob03}}
\label{02fig:14}
\end{figure}
Note that the experimental uncertainty is still statistics
dominated. Thus just running the BNL experiment longer could have
substantially improved the result. Originally the E821 goal was
$\delta a_\mu^\mathrm{exp} \sim 40\times 10^{-11}$.
Fig.~\ref{02fig:amucontrib} illustrates the sensitivity to
various contributions and how it developed in time. The dramatic
\begin{figure}[t]
\vspace*{-3mm}
\centering
\includegraphics[height=7.4cm]{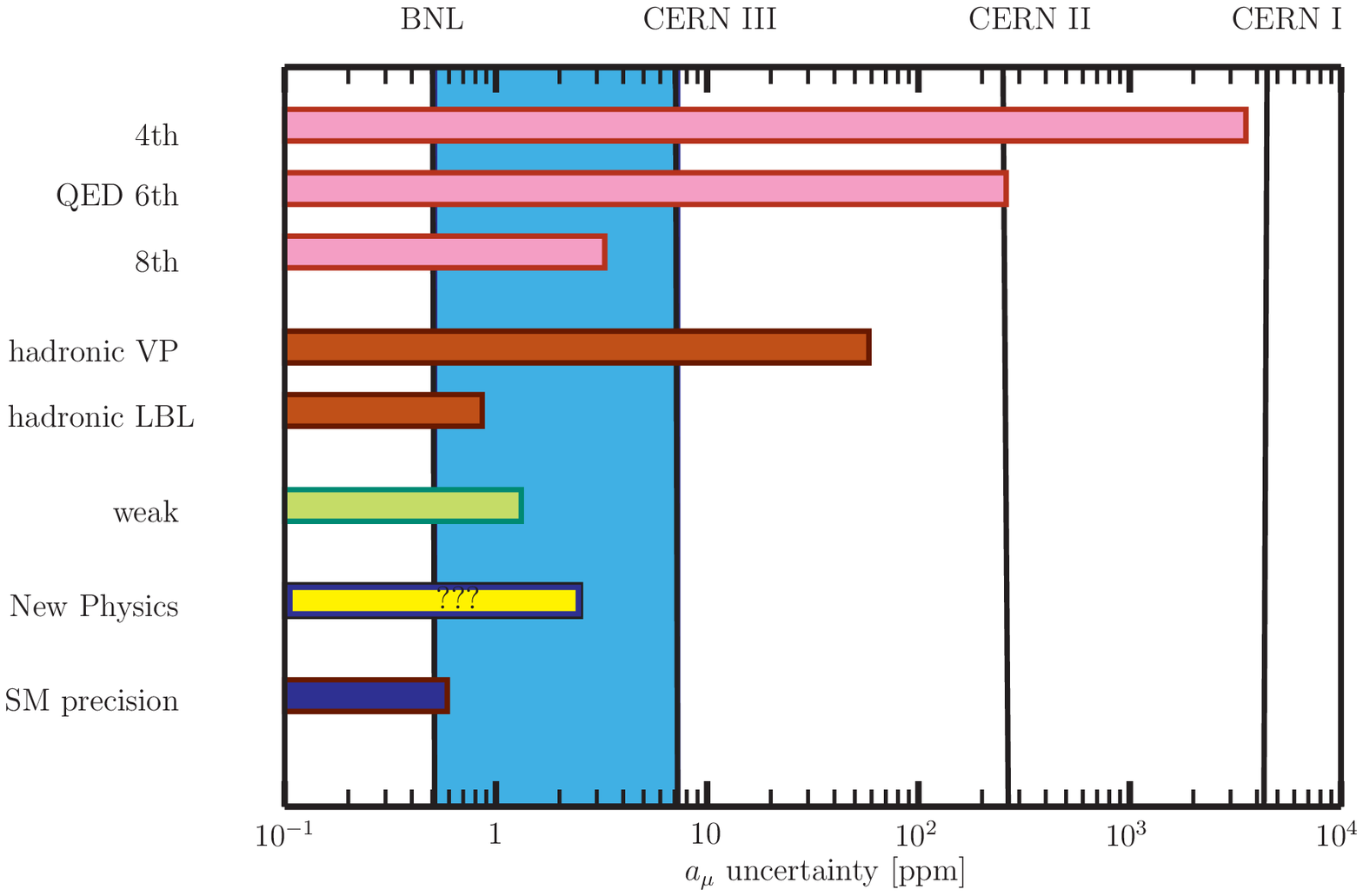}
\caption{Sensitivity of g-2 experiments to various contributions. The
increase in precision with the BNL $g-2$ experiment is shown as a gray
vertical band. New Physics is illustrated by the deviation $(a_\mu^\mathrm{exp}-a_\mu^\mathrm{the})/a_\mu^\mathrm{exp}$}
\label{02fig:amucontrib}
\end{figure}
$(m_\mu/m_e)^2$ enhancement in the sensitivity of $a_\mu$, relative to
$a_e$, to physics at scales $M$ larger than $m_\mu$, which is scaling
like $(m_\mu/M)^2$, and the more than one order of magnitude
improvement of the experimental accuracy has brought many SM effects
into the focus of the interest. Not only are we testing now the
4--loop QED contribution, higher order hadronic VP effects, the
infamous hadronic LbL contribution and the weak loops, we are reaching or
limiting possible New Physics at a level of sensitivity which causes a
lot of excitement. ``New Physics'' is displayed in the figure as the
ppm deviation of
\begin{equation}
\delta a_\mu=a_\mu^\mathrm{exp}-a_\mu^\mathrm{the}=(287\pm91) \times 10^{-11}
\label{02finaldeviat}
\end{equation}
which is $3.2\:\sigma$. We note that the theory error is somewhat
larger than the experimental one. It is fully dominated by the
uncertainty of the hadronic low energy cross--section data, which
determine the hadronic vacuum polarization and, partially, by the
uncertainty of the hadronic light--by--light scattering
contribution.

As we notice, the enhanced sensitivity to ``heavy'' physics is somehow
good news and bad news at the same time: the sensitivity to ``New
Physics'' we are always hunting for at the end is enhanced due to $$a^\mathrm{NP}_\ell
\sim \left(\frac{m_\ell}{M_{\rm NP}} \right)^2$$ by the mentioned mass ratio square,
but at the same time also scale dependent SM effects are dramatically
enhanced, and the hadronic ones are not easy to estimate with the
desired precision.

\section{Prospects}

The BNL muon $g-2$ experiment has determined $a_\mu$ as given by
(\ref{02finalamu}), reaching the impressive precision of 0.54 ppm, a
14--fold improvement over the CERN experiment from 1976. Herewith, a
new quality has been achieved in testing the SM and in limiting
physics beyond it.  The main achievements and problems are
\begin{itemize}
\item a substantial improvement in testing CPT for muons,
\item a first confirmation of the fairly small weak contribution at the
$2-3~\sigma$ level,
\item the hadronic vacuum polarization contribution, obtained via
experimental $e^+e^-$ annihilation data, limits the theoretical precision at the
$1~\sigma$ level,
\item now and for the future the hadronic light-by-light scattering contribution,
which amounts to about $2\sigma$, is not far from being as important
as the weak contribution;  present calculations are model-dependent, and
may become the limiting factor for future progress.
\end{itemize}

At present a $3.2 \sigma$ deviation between theory and experiment is
observed\footnote{It is the largest established deviation between
theory and experiment in electroweak precision physics at present.} and
the ``missing piece'' (\ref{02finaldeviat}) could hint to new physics,
but at the same time rules out big effects predicted by many possible
extensions of the SM.\\

Usually, new physics (NP) contributions are expected to produce
contributions proportional to $m_\mu^2/M_\mathrm{NP}^2$ and thus are
expected to be suppressed by $M_W^2/M_\mathrm{NP}^2$ relative to the
weak contribution.

The most promising theoretical scenarios are supersymmetric (SUSY)
extensions of the SM, in particular the minimal one (MSSM).  Each SM
state $X$ has an associated supersymmetric ``sstate'' $\tilde{X}$
where sfermions are bosons and sbosons are fermions. This implements the
{\em fermion} $\leftrightarrow$ {\em boson} supersymmetry. In addition, an
anomaly free MSSM requires a second complex Higgs doublet, which
means 4 additional scalars and their SUSY partners. Both Higgs fields
exhibit a neutral scalar which aquire vacuum expectation values $v_1$
and $v_2$. Typical supersymmetric contributions to $a_\mu$ stem from
smuon--neutralino and sneutrino-chargino loops
Fig.~\ref{02fig:SUSYgraphs}.
\begin{figure}[t]
\vspace*{-1mm}
\centering
\includegraphics[height=3.2cm]{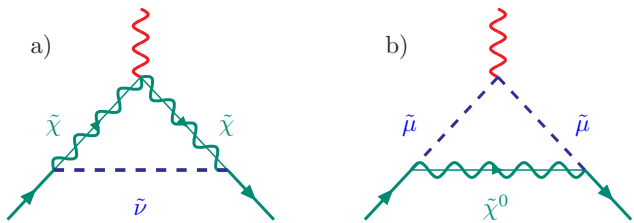}
\caption{Physics beyond the SM: leading SUSY contributions to
$g-2$ in supersymmetric extension of the SM}
\label{02fig:SUSYgraphs}
\end{figure}
Some contributions are enhanced by $\tan\beta\equiv \frac{v_2}{v_1}$
which may be large (in some cases of order { $m_t/m_b\approx
40$}). One obtains~\cite{02Moroi95} (for the extension to 2-loops
see~\cite{02HSW03})
\begin{eqnarray}
a_\mu^{\rm SUSY}
 \simeq \mathrm{sign}(\mu) \frac{\alpha(M_Z)\,(5+\tan^2 \Theta_W)}{ 48\pi\sin^2\Theta_W}\,}{
\frac{m_\mu^2}{\widetilde{m}^2}\:}{ \tan\beta}{ \: \left( 1-\frac{4\alpha}{
\pi}\ln \frac{\widetilde m}{m_\mu}\right)
\end{eqnarray}
$\widetilde{m}=m_{\rm SUSY}$ a typical SUSY loop mass and $\mu$ is the
Higgsino mass term. In the large $\tan\beta$ regime we have
\begin{eqnarray}
\left|a_\mu^{\rm SUSY}\right|
 \simeq 123\times 10^{-11} \left( \frac{100 {\rm\ GeV}}{
\widetilde{m}}\right)^2 \tan\beta \;.
\end{eqnarray}
$a_\mu^{\rm SUSY}$ generally has the same sign as the $\mu$-parameter.
The deviation (\ref{02finaldeviat}) requires positive $\mathrm{sign}(\mu)$
and if identified as a SUSY contribution
\begin{eqnarray}
\widetilde m \simeq (65.5\mbox{ GeV})\sqrt{\tan\beta} \;.
\end{eqnarray}
Negative $\mu$ models give the opposite sign contribution to $a_\mu$ and
are strongly disfavored.  For $\tan\beta$ in the range $2 \div 40$ one obtains
\begin{eqnarray}
\widetilde m \simeq 92-414 \mbox{ GeV} \;,
\end{eqnarray}
precisely the range where SUSY particles are often expected.  For a
variety of non-SUSY extensions of the SM typically $\left|
a_\mu(\mbox{NP}) \right| \simeq {\cal C}\:m_\mu^2/M^2$ where ${\cal C}
=O(1)$ [or $O(\alpha/\pi)$ if radiatively induced].  The current
constraint suggests (very roughly) $M\simeq 1.7-2.4 \mbox{ TeV}$
[$M\simeq 87-121 \mbox{ GeV}$]. The ${\cal C} =O(1)$ assumption is
problematic, however, since no tree level contribution can be
tolerated. For a more elaborate discussion and further references I
refer to~\cite{02CzMa01}. Note that the most natural leading
contributions in extensions of the SM are 1-loop contributions similar
to the leading weak effects or the leading MSSM
contributions. However, mass limits set by LEP and Tevatron make it
highly non-trivial to reconcile the observed deviation to many of the
new physics scenarios.  Only the $\tan \beta$ enhanced contributions
in SUSY extensions of the SM for $\mu >0$ and large enough $\tan
\beta$ may explain the ``missing contribution''. Two Higgs doublet
models~\cite{02Krawczyk05} have similar possibilities. Physics beyond
the SM of course not only contributes to $a_\mu$ but also to other
observables like to the branching fraction $Br(b \to s
\gamma)=(3.40\pm0.28)\times 10^{-4}$ or to the $W$ mass prediction
$M_W=80.392(29)~\mathrm{GeV}$. In the $R$--parity conserving MSSM
the lightest neutralino is stable and therefore is a candidate for
cold dark matter in the universe. From the precision mapping of the
anisotropies in the cosmic microwave background, the WMAP
collaboration has determined the relict density of cold dark matter
to $\Omega {\rm h}^2=0.1126\pm0.0081$. This sets severe constraints on
the SUSY parameter space (see for example~\cite{02HBaeretal04}).

Of course, for a specific model, one must check that the sign of
the induced $a_\mu^{\rm NP}$}is in accord with experiment
(i.e. it should be positive).

Plans for a new $g-2$ experiment exist~\cite{02LeeRob03}. In fact, the
impressive 0.54 ppm precision measurement by the
E821collaboration at Brookhaven was still limited by statistical
errors rather than by systematic ones. Therefore an upgrade of the
experiment at Brookhaven or J-PARC (Japan) is supposed to be able to
reach a precision of 0.2 ppm (Brookhaven) or 0.1 ppm (J-PARC).

For the theory this poses a new challenge.  It is clear that on the
theory side, a reduction of the leading hadronic uncertainty is
required, which actually represents a big experimental challenge: one
has to attempt cross-section measurements at the 1\% level up to
$J/\psi$[$\Upsilon$] energies (5[10] GeV). Such measurements would be
crucial for the muon $g-2$ as well as for a more precise determination
of the running fine structure constant $\alpha_{\rm QED}(E)$. In
particular, $e^+e^-$ low energy cross section measurements in the
region between 1 and 2.5 GeV~\cite{02VEPP2000,02KLOE2} are able to
substantially improve the accuracy of $a_\mu^\mathrm{had(1)}$ and
$\alpha_{\rm QED}(M_Z)$~\cite{02FJ06}.

New ideas are required to get less model--dependent estimations of the
hadronic LbL contribution. Here, new high statistics experiments
attempting to measure the $\pi^0\gamma^*\gamma^*$ form factor ${\cal
F}(m_\pi^2,-Q_1^2,-Q_2^2)$ for $Q_1^2 \sim Q_2^2$ and a scan of the
light-by-light off-shell amplitude via $e^+e^- \to e^+ e^- \gamma^*
\gamma^* \to e^+e^- \gamma \gamma$ would be of great
help.  Certainly lattice QCD studies~\cite{02BluAub06} will be able to shed
light on these non-perturbative problems in future.

In any case the muon $g-2$ story is a beautiful example which
illustrates the experience that \textit{the closer we look the more there is
to see}, but also the more difficult it gets to predict and interprete
what we see.  Even facing problems to pin down precisely the hadronic
effects, the achievements in the muon $g-2$ is a big triumph of
science. Here all kinds of physics meet in one single number which is the
result of a truly ingenious experiment. Only getting all details in
all aspects correct makes this number a key quantity for testing our
present theoretical framework in full depth. It is the result of
tremendous efforts in theory and experiment and on the theory side has
contributed a lot to push the development of new methods and tools
such as computer algebra as well as high precision numerical methods
which are indispensable to handle the complexity of hundreds to
thousands of high dimensional integrals over singular integrands
suffering from huge cancellations of huge numbers of
terms. Astonishing that all this really works!\\[3mm]

Note added: After completion of this work a longer review
article appeared~\cite{MdeRLR07},  which especially reviews the
experimental aspects in much more depth than the present essay. For a
recent reanalysis of the light-by-light contribution we refer the
reader to~\cite{BijnensPrades07}, which presents the new estimate
$a_\mu^{\rm LbL}=(110\pm40)\times 10^{-11}$.\\[3mm]

\noindent
{\bf Acknowledgments} \\

This extended update and overview was initiated by a talk given at the
International Workshop on Precision Physics of Simple Atomic Systems
(PSAS 2006). It is a pleasure to thank the organizers and in
particular to Savely Karshenboim for the kind invitation to this
stimulating meeting. The main new results were first presented at the
Kazimirez Final EURIDICE Meeting. Thanks to Maria Krawczyk and Henryk
H.~Czy\.z for the kind hospitality in Kazimirez. Particular thanks to
Andreas Nyffeler and to Simon Eidelman for many enlightening
discussions. Thanks also to Oleg Tarasov and Rainer Sommer for helpful
discussions and for carefully reading the manuscript. Many thanks to
B. Lee Roberts and the members of the E821 collaboration for many
stimulating discussions over the years and for providing me some of
the figures. Special thanks go to Wolfgang Kluge, Klaus M\"onig,
Stefan M\"uller, Federico Nguyen, Giulia Pancheri and Graziano
Venanzoni for numerous stimulating discussions and their continuous
interest. I gratefully acknowledge the kind hospitality extended to me
by Frascati National Laboratory and the KLOE group.  This work was
supported in part by EC-Contracts HPRN-CT-2002-00311 (EURIDICE) and
RII3-CT-2004-506078 (TARI).

\end{document}